\documentclass[a4paper,fleqn,usenatbib]{mnras}
\usepackage{newtxtext,newtxmath}

\usepackage[T1]{fontenc}
\usepackage{ae,aecompl}

\usepackage{graphicx}	
\usepackage{amsmath}	
\usepackage{amssymb}	

\title[Orbital structure of A\,85]{Deep spectroscopy in nearby galaxy clusters: III Orbital structure of galaxies in Abell 85}

\author[J. A. L. Aguerri et al.]{J. A. L. Aguerri$^{1,2}$\thanks{jalfonso@iac.es}, I. Agulli$^{1,2,3,4}$\thanks{ireagu@iac.es}, A. Diaferio$^{3,4}$, C. Dalla Vecchia$^{1,2}$ \\
$^{1}$ Instituto de Astrof\'{\i}sica de Canarias, C/ V\'{\i}a L\'actea s/n, E-38205, La Laguna, Spain\\
$^{2}$ Departamento de Astrof\'{\i}sica, Universidad de La Laguna, E-38206, La Laguna, Spain\\
$^{3}$ Dipartimento di Fisica, Universit\`a di Torino, Via P. Giuria 1, I-10125 Torino, Italy \\
$^{4}$ Istituto Nazionale di Fisica Nucleare (INFN), sezione di Torino, Via P. Giuria 1, I-10125 Torino, Italy
}

\date{Accepted XXX. Received YYY; in original form ZZZ}

\pubyear{2016}

\begin{document}
\label{firstpage}
\pagerange{\pageref{firstpage}--\pageref{lastpage}}
\maketitle

\begin{abstract}
Galaxies in clusters are strongly affected by their environment. They evolve according to several physical mechanisms that are active in clusters. Their efficiency can strongly depend on the orbital configuration of the galaxies.
Our aim is to analyse the orbits of the galaxies in the cluster Abell 85, based on the study of the galaxy velocity anisotropy parameter.
   We have solved the Jeans equation under the assumption that the galaxies in A\,85 are collisionless objects, within the spherically symmetric gravitational potential of the virialized cluster. The mass of the cluster was estimated with X-ray and caustic analyses. 
   We find that the anisotropy profile of the full galaxy population in A\,85 is an increasing monotonic function of the distance from the cluster centre: on average, galaxies in the central region $(r/r_{200} < 0.3$) are on isotropic orbits, while galaxies in the outer regions are on radial orbits. We also find that the orbital properties of the galaxies strongly depend on their  stellar colour. In particular, blue galaxies are on less radial orbits than red galaxies.  The different families of cluster galaxies considered here have the pseudo phase-space density profiles $Q(r)$ and $Q_{r}(r)$ consistent with the profiles expected in virialized dark matter halos in $N$-body simulations. This result suggests that the galaxies in A\,85 have reached dynamical equilibrium within the cluster potential. 
   Our results indicate that the origin of the blue and red colour of the different galaxy populations is the different orbital shape rather than the accretion time.
\end{abstract}



\section{Introduction}

 Collisionless and virialized systems obey the Jeans equation which combines the gravitational potential  with the density and the velocity distribution of tracers that move under that potential. The collisionless nature of galaxies in clusters allows the use of Jeans analysis to infer the total mass and dark matter distribution of clusters from the measurement of the projected number density of galaxies  and their line-of-sight velocity, under the assumption of virial equilibrium and spherical symmetry. However, it is also necessary to assume the orbital distribution, namely the velocity anisotropy parameter $\beta$, of the galaxies in the cluster. This assumption generates the so-called "mass-orbit" degeneracy \citep[see e.g.,][]{merritt1987}. 
 
 To break this degeneracy, we can infer the orbital distribution of cluster galaxies by using independent measures of the cluster mass profile, based on X-ray observations \citep[see e.g.][]{vikhlinin2006,schmidt2007}, galaxy velocity dispersion \citep[e.g.][]{diaferio2005,rines2006}, Sunyaev-Zeldovich measurements \citep[see][]{atrio2008}, or weak and strong gravitational lensing studies \citep[see][]{umetsu2010}. With this approach, the Jeans analysis only provides the orbital anisotropy profile of the galaxies in the cluster. This profile gives us information about the orbital configuration of the galaxies, indicating whether the galaxies are moving on isotropic, radial or tangential orbits.  This piece of information is important to have a global picture about the effects of the environment on the evolution of galaxies. 
 
 There are several mechanisms affecting  galaxies in clusters and not in the field. In particular, galaxies in high density environments suffer from gravitational and hydrodynamical effects \citep[see e.g.][ for a review]{blanton2009}. Galaxies lose dark matter and baryonic mass due to tidal interactions with the global cluster potential and in high speed encounters with other cluster members, the so-called galaxy harassment \citep[see][]{moore1998}.  Most of this mass loss is due to interactions between the galaxy and the cluster potential; only about 30$\%$  is due to high speed encounters with other cluster members \citep[see][]{gnedin2003,knebe2006,smith2010}. In addition, harassment is more efficient on low-mass and low-surface brightness cluster members \citep[][]{mastropietro2005,aguerri2009,smith2010}.  Besides these tidal effects, galaxies passing near the cluster centre also suffer from ram pressure stripping \citep[see][]{gunn1972,quilis2000}. This effect produces a swept of the cold gas of the galaxies on a short time-scale. The supply of gas into the galaxies can be halted by the so-called strangulation mechanism \citep[see ][]{larson1980, fujita2004}. These tidal and hydrodynamical effects are responsibles for the observational properties of cluster galaxies \citep[see][]{blanton2009}. Numerical simulations show that the mass loss of the galaxies in clusters strongly depends on their orbits and orientation \citep[see][]{smith2010, smith2015, bialas2015}.  Thus, galaxies on eccentric orbits with small pericentre and those on circular orbits with small radius could be strongly disturbed by the gravitational potential and suffer important transformations. Therefore, the study of the orbital distribution of galaxies in clusters is crucial in order to understand the evolution of galaxies in high density environments.
  
 Numerical simulations show that the orbital anisotropy profile of dark matter sub-halos in cluster-size halos tipically increases with radius \citep[see e.g.][]{lemze2012}. This behaviour is common to systems formed by gravitational collapse. Nevertheless, there is a large variety of anisotropy profiles among different clusters. In addition, numerical simulations suggest that the anisotropy parameter is different for different galaxy populations and evolves with time \citep[see][]{iannuzzi2012}. In particular, these simulations show that blue galaxies have smaller anisotropy parameters than red galaxies at all redshifts.  Indeed, the  blue galaxies observed in the cluster at the present epoch are those entering  the cluster potential with the most tangential orbits.

  Several studies in the past have studied the orbits of galaxies in clusters, with some focusing  on invidual clusters \citep[see][]{kent1982,hwang2008,wojtak2010,biviano2013,munari2014,annunziatella2014}. A large number of redshifts is desirable in order to measure the radial anisotropy profile with small uncertainties. In this case, studies on  stacked clusters were performed based on large spectroscopic surveys such as CNOC \citep[see][]{carlberg1997}, CAIRNS \citep[see][]{rines2003}, or ENACS \citep[see][]{biviano2004}.  Studies on individual clusters reported  a variety of orbits of galaxies. The pioneering work by \cite{kent1982} found that the orbits of the galaxies in the Coma cluster are not mainly radial. A significant fraction of galaxies should be on tangential orbits. Studies on stacked clusters from the CNOC and CAIRNS surveys showed that orbits of galaxies in clusters are consistent with being isotropic or moderately radial. This result was also found by \cite{hwang2008} for a large sample of clusters using spectroscopic data from the SDSS and 2dFGRS surveys. \cite{biviano2004} found that the orbital distribution of bright ellipticals, early and late-type spirals were different, with early-type spirals on isotropic orbits, and isotropic orbits statistically rejected for late-type spirals at large radii. However, other studies, such as \cite{hwang2008}, did not find this difference between early and late-type galaxies. 
  
Previous studies focused on the bright galaxy population. Little is reported about the orbital structure of dwarf galaxies in clusters. \cite{adami2009} found that Coma dwarf galaxies have radially anisotropic orbits even close to the cluster centre. They concluded that dwarf galaxies in Coma could be the remnants of galaxies that fell into the cluster on radial orbits. The study made by \cite{annunziatella2016} on the intermediate redshift cluster Abell 209 (A\ 209) found that low-mass passive galaxies have radial orbits, similar to massive galaxies at large clustercentric distances. However, they do have tangential orbits in the cluster centre and avoid small pericenters close to the brightest cluster galaxy. They concluded that the dwarf galaxies observed today in A\ 209 are those which were not destroyed near the cluster centre because they moved on tangential orbits. The extension of this kind of studies to a broader range of environments remains to be done and will provide observational constraints on the origin and evolution of low mass halos in galaxy clusters. Moreover, the large variety of velocity anisotropy profiles among clusters and the different results found by numerical simulations and observations point towards the importance of the analysis of the orbital structure of individual clusters to have a global picture of galaxy evolution.

In the present work, we will analyse the orbital structure of the galaxies in the nearby and massive cluster Abell 85 (hereafter A\,85). Our deep spectroscopy and the richness of this cluster enable us to analyse its orbital structure  down to  the dwarf regime, with absolute magnitude $\sim M^{*}_{r} + 6$. This paper is the third of a series analysing the properties of the galaxy population in A\,85. In the two previous papers, we analysed the galaxy luminosity function (LF) and its dependence on both the cluster environment and the galaxy population properties \citep[see][]{agulli2014, agulli2016}. We found that the faint-end slope of the LF of A\,85 is similar to the slope of the field LF. We found a difference in the nature of the low-mass galaxy populations: red dwarfs dominate the faint-end of the LF in the cluster, while blue dwarfs dominate the field population. Our result suggests that the environment quenches the star formation and turns the stellar population red in low-mass galaxies \citep[see][]{agulli2014}. We also analysed the variation of the LF as a function of environment in A\,85. The bright ($M_{r} < -21.5$) and faint ($M_{r} > -18.0$) ends of the LF do not depend on the cluster environment. In contrast, the intermediate luminosity range of the LF $(-21.5 < M_{r} < -18.0)$ shows a dependence on the cluster position. These findings were interpreted as a consequence of the dynamical friction suffered by the galaxies in clusters and of the switching off of star formation due to hydrodynamical processes \citep[see][]{agulli2016}. The present work will complete this picture by estimating the orbital anisotropy of the full population of cluster members and of different subsets selected by luminosity or stellar colour. 
 
 The paper is organised as follows. Section 2 describes the data and the computational procedure followed for obtaining the radial anisotropy profile of the galaxies. Section 3 shows the results. The discussion and conclusions are given in Sections 4 and 5, respectively. Throughout this paper we  use the cosmological parameters $\Omega_{\Lambda} = 0.7$, $\Omega_{m}=0.3$, and $H_{0}=75$ Mpc$^{-1}$ km s$^{-1}$.


\section{Orbital structure of galaxy clusters: the case of Abell\ 85}

A collisionless and spherically symmetric galaxy cluster  in dynamical equilibrium obeys the Jeans equation given by:
\begin{equation}
\frac{d}{dr}[\nu_{g}(r)\sigma_{r}^{2}(r)] + \frac{2 \beta(r)}{r} \nu_{g}(r) \sigma_{r}^{2}(r) = -\nu_{g}(r) \frac{GM(r)}{r^{2}},
\label{eq1}
\end{equation}
where $G$ is the gravitational constant, $M(r)$ is the total gravitational mass of the system contained in a sphere of radius $r$, $\nu_{g}(r)$ is the number density of cluster galaxies at radius $r$, $\sigma_{r}(r)$ is the radial component of the velocity dispersion, and $\beta(r)= 1 - \sigma_{\theta}^{2}(r)/\sigma_{r}^{2}(r)$ is the velocity anisotropy parameter. The parameter $\sigma_{\theta}(r)$ is the tangential velocity dispersion. In the absence of rotation, $\sigma_{\theta}(r)$ is equal to the azimuthal velocity dispersion $\sigma_{\phi}(r)$. According to this definition, clusters with $\beta = 0$ correspond to systems with an isotropic velocity distribution, clusters with $\beta = 1$ have galaxies on fully radial orbits, and $\beta \rightarrow \infty$ indicate fully tangential orbits.

Equation \ref{eq1} shows that $\beta(r)$ can be estimated when $\nu_{g}(r)$, $\sigma_{r}^{2}(r)$ and $M(r)$ are known. However, in observational astronomy, only the projections of some quantities can be measured. In particular, we can measure $\Sigma(R)$, $\sigma_{p}(R)$ and $M(r)$, with $\Sigma(R)$ the projected number density of galaxies at projected distance $R$ from the cluster centre, and $\sigma_{p}$  the line-of-sight velocity dispersion. It was probed  that it is possible to derive $\beta(r)$ when $\Sigma(R)$, $\sigma_{p}(R)$ and $M(r)$ are known, by using the S$^2$ method, first proposed by \cite{binney1982} and used in various works related to galaxy clusters \citep[see e.g.][]{solanes1990,biviano2004,katgert2004,adami2009,mamon2013}.

We applied this method in order to obtain the orbital structure of the galaxies in the nearby galaxy cluster A\,85.

\subsection{The spectroscopic data of A\,85}

The cluster A\,85 is a nearby  ($z=0.055$) and massive system ($r_{200}=1.02 h^{-1}$ Mpc; $M_{200} = 2.5 \times 10^{14} h^{-1} M_{\sun}$ \cite{rines2006}). It is one of the brightest X-ray clusters located in the southern hemisphere. It has been extensively studied over a large range of wavelengths. We can mention observations of this cluster in radio \citep[][]{slee2001} and optical wavelengths \citep[][]{rines2006, bravoalfaro2009,aguerri2007,agulli2014}, and in X-ray \citep[][]{ramella2007,schenck2014,ichinohe2015}. It is well known that this cluster is not completely virialized. Substructures can be found in both the galaxy distribution and the X-ray maps \citep[see][]{durret2003,ramella2007,boue2008,aguerri2010,schenck2014, yu2016}.  

\begin{figure}
   \centering
\includegraphics[width=\hsize]{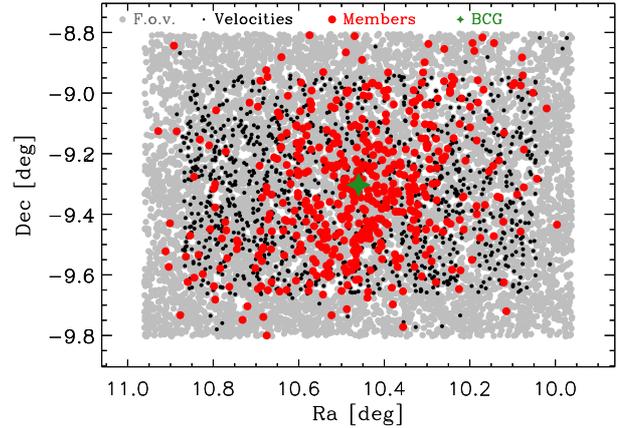}
   \caption{Spatial distribution of galaxies in the direction of A\,85. The targets for the spectroscopic observations are shown in grey colour. They correspond to objects classified as galaxies by the SDSS-DR6 and with stellar colour $g - r = 1.0$. The black dots indicate the galaxies with measured redshift. The red dots correspond to galaxies  identified as cluster members. The location of the brightest cluster galaxy is indicated by a green star.}
             \label{members}%
    \end{figure}

We carried out an extensive spectroscopic survey  of A\,85. The observations cover an area of 3.0 $\times$ 2.6 Mpc$^{2}$ around the cluster centre ($\alpha$ (J2000): 00$^{h}$ 41$^{m}$ 5.448$^{s}$ $\delta$ (J2000): -9$^{o}$ 18$^{'}$ 11.45$^{''}$). The spectroscopy was performed with the VIsible Multi-Object Spectrograph at the Very Large Telescope (VIMOS@VLT) in combination with AutoFib2/WYFFOS at the William Herschel Telescope (WHT). The large capacity to take spectra  of both instruments and the data already available from the literature have allowed us to collect 1603 redshifts in the field of A\,85 spanning an apparent magnitude interval $13 < m_{r} < 22$, within 1.4 $r_{200}$, and for galaxies bluer than $g-r = 1.0$. Figure \ref{members} shows the targets, measured velocities and cluster members obtained by using this dataset. An extensive presentation of the data and the data reduction process was given in  \cite{agulli2016}.

Our catalogue, together with literature spectroscopic redshifts, was analysed by using the caustic method \citep[][]{diaferio1997,diaferio1999,serra2011}, that returns a list of cluster members where, on average, only 2\% of the galaxies within  $r_{200}$ are interlopers \citep{serra2013}. We identified 460 galaxy members within 1.4 $r_{200}$ down to $M_{r} \sim -16.0$. This sample is ideal to perform studies of galaxy orbits in individual clusters. In particular, the large number of cluster members together to the homogeneous spatial distribution of the redshifts allow us to analyse the orbital structure of different families of galaxies selected by different physical properties, such as luminosity and stellar colour.

\subsection{Virialized galaxies}

One of the assumptions of the Jeans equation is the dynamical equilibrium of the system. However, due to the continuous accretion of smaller sub-units, clusters are never  in perfect equilibrium.  

There are many statistical tests developed to detect galaxy substructures in clusters \citep[see e.g.][ and references therein]{aguerri2010,zarattini2015}. It is well known that  A\,85 has galaxies in substructures \citep[see e.g. ][]{aguerri2010,ramella2007}. Some of the substructures are also visible in the X-ray surface brightness or temperature distribution \citep[see][]{ramella2007,schenck2014,ichinohe2015, yu2016}. In order to apply Eq. \ref{eq1} to determine $\beta(r)$, we only have to consider those galaxies in dynamical equilibrium within the gravitation potential well of A\,85.  In this work we used  the caustic method \citep[see][]{diaferio1999,serra2011,yu2015} which provides an accurate identification of substructures \citep[see][]{yu2015}. We counted 102 galaxies in substructures, representing about $ 20 \%$ of the cluster members. Only the remaining 358 galaxy members were considered for the determination of $\beta(r)$. For an extensive explanation of the substructure determination in A\,85, we refer  the reader to \cite{agulli2016} and \cite{yu2016} 

\subsection{The galaxy number density profile}

The projected galaxy number density profile, $\Sigma(R)$, is the number of galaxies per Mpc$^{2}$ as a function of the projected radius, $R$. Assuming a spherically symmetric cluster, this observed quantity is related to the galaxy number density profile, $\nu_{g}(r)$, by
\begin{equation}
\nu_{g}(r)=-\frac{1}{\pi}\int_{r}^{\infty} \frac{d\Sigma}{dR}\frac{dR}{\sqrt{R^{2}-r^{2}}}\; .
\label{eqnu}
\end{equation}
This equation is known as the inverse Abel transform and is used in several astronomical problems related to galaxies and clusters \citep[see e.g.][]{stark1977,simonneau1998,aguerri2001,trujillo2002,biviano2004,mendezabreu2008}.

\begin{figure*}
   \centering
\includegraphics{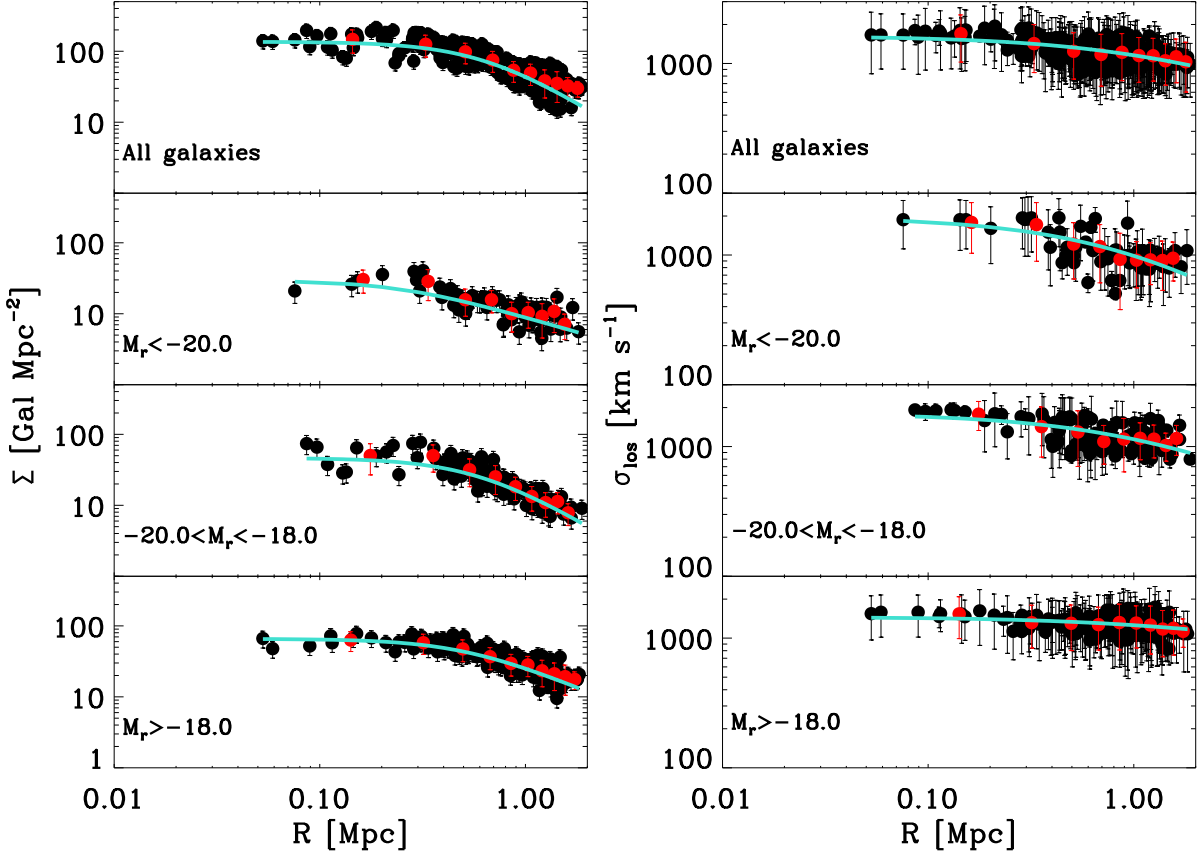}
   \caption{Projected galaxy number density profiles (left-hand panels) and projected velocity dispersion profiles (right-hand panels). From top to bottom we show the profiles of all the galaxies and of the galaxy sub-samples using different magnitude cuts. The black points represent the projected galaxy number density and the projected velocity dispersion at each galaxy position. The solid cyan lines are the smoothed fits of the quantities (see text for more details). The red points represent the projected galaxy number density and projected velocity dispersion obtained in radial bins.}
             \label{densisigma}%
    \end{figure*}

We computed $\Sigma(R)$ at the position of every  member of A\,85 by taking into account the $\sqrt{N_{m}}$ nearest cluster members \citep[see][]{katgert2004}. The quantity $N_{m}$ is the number of members not in substructures. Border effects were taken into account in the computation. Values of $N_m$ in the range of 10 - 20 nearest neighbors do not affect our final results.  The left panels of Fig. \ref{densisigma} show  $\Sigma(R)$ for all the galaxies and for the galaxies in three different magnitude bins. As a comparison, we overplot  the values of $\Sigma(R)$ obtained in radial bins. The completeness of the spectroscopic data does have a mild radial variation with a mean value of $\sim 60\%$. Therefore we did not apply any incompleteness correction to the calculation of $\Sigma (R)$. We do not expect radial variation for the completeness of the spectroscopic data for the different subsamples of galaxies considered in this work. The spectroscopic targets of these subsamples were selected without any radial trend, similarly to the total sample.

In order to apply the Jeans equation and  derive the velocity anisotropy profile, we need smoothed estimates of $\Sigma(R)$ \citep[see e.g][]{biviano2004}. In this work,  we fitted the observations with the analytical function
\begin{equation}
\Sigma_{s}(R) = \Sigma_{0} [1 + (R/R_{c})^{2})]^{\beta_{m}}
\end{equation}
where $\Sigma_0$ is the central value of the projected number density, $R_{c}$ is the core radius, and $\beta_{m}$ is the power law exponent. The smoothed projected galaxy density profile was used in Eq. \ref{eqnu}.  Table \ref{tabgalden} shows the parameters of  the fits to the galaxy density profiles for the different galaxy samples considered.

In the left panel of Fig. \ref{densisigma}, we show  the galaxy number density computed in radial bins. The fitted profiles are good representation of the binned data as confirmed by the unreduced $\chi^{2}$ given in Table \ref{tabgalden}.

\begin{table}
\caption{Parameters of the fits to the galaxy number density profiles for the different galaxy samples}            
\label{tabgalden}  
\centering        
\begin{tabular}{c c c c c c}        
\hline\hline                
Galaxies & $\Sigma_{0}$ & $R_{c}$ & $\beta_{m}$ & $\chi^{2}$ & DOF \\ 
           &    [Gal Mpc$^{-2}$]    &     [Mpc]   &                   \\  
\hline   
   All & 162 & 0.6 & -0.9 & 1.9 & 7 \\      
   All red & 151 & 0.5 & -0.9 & 1.9 & 7 \\
   All Blue & 19 & 0.5 & -0.9 & 1.3 & 6 \\
   $M_{r} \le -20.0$ & 31 & 0.2 & -0.4 & 1.6 & 6 \\
   $- 20.0 \le M_{r} \le -18.0$ & 47 & 0.6 & -0.9 & 1.4 & 6 \\
   $M_{r} \ge -18.0$ & 66 & 0.5 & -0.6 & 1.2 & 6 \\
   Dwarf red & 64 & 0.4 & -0.6 & 0.6 & 5 \\
   Dwarf blue & 14 & 0.9 & -0.3 & 1.0 & 5 \\
\hline                                
\end{tabular}
Note: $\chi^{2}$ represents the goodness between the the fitted galaxy number density profiles (cyan lines in Fig. \ref{densisigma}) and the binned data (red circles in Fig. \ref{densisigma}). DOF shows the degrees of freedoms
\end{table}

\subsection{The galaxy velocity dispersion profile}

The procedure to determine the projected velocity dispersion profile, $\sigma_{p} (R)$, is similar to the one for the number density profile described above. We estimated $\sigma_{p}(R)$ at the position of each member by taking into account the $\sqrt{N_{m}}$  closest neighbours, then we obtained a smoothed version of the profile by fitting a simple power law \citep[see e.g.][]{carlberg1997}
\begin{equation}
\sigma_{p,s}(R)=\sigma_{p,0} [1+ R]^{p}
\end{equation}
where $\sigma_{p,0}$ is the central projected velocity dispersion and $p$ is the exponent. The right panels of Fig. \ref{densisigma} show the line-of-sight velocity dispersion profiles for the full sample of galaxies in A\,85 and different families of galaxies selected by their luminosity. Table \ref{tabgalsig} shows the parameters of the fits to the galaxy velocity dispersion profiles for the different galaxy samples considered.

We also overplotted the $\sigma_{p}(R)$ computed in radial bins. Note the similarities between $\sigma_{p}$ measured in radial bins and by using local cluster members (see the unreduced $\chi^{2}$ values given in Table \ref{tabgalsig}). 

\begin{table}
\caption{Parameters of the fits to the galaxy velocity dispersion profiles for the different galaxy samples}            
\label{tabgalsig}  
\centering        
\begin{tabular}{c c c c c c}        
\hline\hline                
Galaxies & $\sigma_{p,0}$  & $p$ & $\chi^{2}$ & DOF & \\ 
           &    [km s$^{-1}$]      &       &  &    &        \\  
\hline   
   All & 1326 & -0.3 & 0.9 & 8 & \\      
   All red & 1530 & -0.7 & 1.8 & 8 & \\
   All Blue &  2144 & -0.4 & 0.6 & 7 &\\
   $M_{r} \le -20.0$ &  1160 & -0.4 & 2.1 & 7 &\\
   $- 20.0 \le M_{r} \le -18.0$ & 1395 & -0.4 & 2.1 & 7 & \\
   $M_{r} \ge -18.0$ &  1438 & -0.3 & 2.1 & 7 &\\
   Dwarf red &  1319 & -0.4 & 0.8 & 6 &\\
   Dwarf blue &  1618 & -0.1 & 0.3 & 6 & \\
\hline                                
\end{tabular}
Note: $\chi^{2}$ represents the goodness between the the fitted velocity dispersion profiles (cyan lines in Fig. \ref{densisigma}) and the binned data (red circles in Fig. \ref{densisigma}). DOF shows the degrees of freedoms
\end{table}

\subsection{The mass profile}

The halo mass distribution of A\,85 has been investigated in several works   \citep[see][]{reiprich2002,sandersen2003,demarco2003,durret2005,rines2006}. In the present work, we considered those models assuming a Navarro-Frenk-White (NFW) mass density profile \citep[][]{navarro1997} 
\begin{equation}
\rho(r)=\frac{3 H_{0}^{2}}{8 \pi G} \frac{\delta_{c}}{c (r/r_{s}) (1+cr/r_{s})^{2}}
\end{equation}
where $r_{s}$ is a scale radius, $\delta_{c}$ is a characteristic density, $c=r_{200}/r_{s}$ is the concentration parameter and $\delta_{c}=\frac{200}{3}\frac{c^{3}}{[ln(1+c) -c/(1+c)]}$. The corresponding mass profile is 
\begin{equation}
M(r) = \frac{3}{2} \frac{H_{0}^{2}}{G} \delta_{c} r_{s}^{3} [ln (\frac{r_{s}+r}{r_{s}}) - \frac{r}{r_{s}+r}]\, .
\end{equation}
We computed the profile of the velocity anisotropy parameter of A\,85 by using the NFW mass profiles obtained by \cite{sandersen2003,durret2005,rines2006}. Table \ref{tabnfw} shows the concentration parameter and the $r_{200}$ radius of these mass models. The considered mass profiles were obtained from different data sets: \cite{sandersen2003} and \cite{durret2005} obtained the mass profile 
 from X-ray data, while \cite{rines2006} used the caustic method applied to spectroscopic galaxy redshifts. 

Given the discrepancies between the different halo mass determinations, we defined a mean mass distribution given by a NFW profile with  $c= 4.0$ and $r_{200} = 1.44 $ Mpc. These parameters correspond to the mean values of the three mass models considered.  The uncertainty in the mass estimate is then included in the determination of the anisotropy. 


We did not use the mass profile estimated with the caustic method applied to our data because our data only cover the area within 1.4 $r_{200}$. Indeed, the  technique, on average, returns a biased mass estimate within $r_{200}$ \citep[see][]{diaferio1999,serra2011}. However, the caustic method on average returns unbiased NFW fit parameters when the fit is performed to large radii \citep{serra2011}.

\begin{table}
\caption{Parameters of the NFW profiles of A\,85}            
\label{tabnfw}  
\centering        
\begin{tabular}{c c c c}        
\hline\hline                
Model & $c$ & $r_{200}$ & Reference \\ 
           &        &     (Mpc)   &                   \\  
\hline   
   NFW1 & 2.4 & $1.68$ & \cite{sandersen2003} \\      
   NFW2 & 5.0 & 1.30    & \cite{durret2005} \\
   NFW3 & 4.5 & 1.33     & \cite{rines2006} \\
\hline                                
\end{tabular}
\end{table}

\subsection{The orbital anisotropy parameter}

\cite{binney1982} was the first to show that $\beta$ can be obtained when $\Sigma(R)$, $\sigma_{p}(R)$ and $M(r)$ are known \citep[for an application of this method see e.g.][]{biviano2004}. 

The observed profiles $\Sigma(R)$ and $\sigma_{p}(R)$ are used to define the function 
\begin{equation}
H(R)=\frac{1}{2} \Sigma(R) \sigma_{p}^{2}(R)\; .
\label{eqh}
\end{equation}
This function is used to calculate $K(r)$, another function given by the Abel integral
\begin{equation}
K(r)= 2 \int^{\infty}_{r} H(x) \frac{xdx}{\sqrt{x^{2}-r^{2}}}\; .
\label{eqk}
\end{equation}
In addition, the estimates of the mass profile, $M(r)$, and of the number galaxy density, $\nu_{g}$, are used to compute the function 
\begin{equation}
\Psi(r) = -G M(r) \nu_{g}(r)/r^{2}\; .
\end{equation}

Equations \ref{eqh} and \ref{eqk} can be used to derive two equations for $\sigma_{r}(r)$ and $\beta(r)$
\begin{eqnarray}
[3-2\beta(r)] \sigma_{r}^{2}(r) = \frac{-1}{\nu(r)} \int_{r}^{\infty} \Psi(r) dx - \frac{2}{\pi r \nu(r)}\frac{dK(r)}{dr}
\label{eqsis1}
\end{eqnarray}

\begin{equation}
\begin{array}{r}
\beta(r) \sigma_{r}^{2}(r) = \frac{1}{\nu(r) r^{3}} \int_{0}^{r} x^{3} \Psi(x) dx 
+ \frac{1}{\pi r \nu(r)} \frac{dK(r)}{dr} - \frac{3 K(r)}{\pi r^{2} \nu(r)} + \\
+ \frac{3}{\pi r^{3} \nu(r)} \int_{0}^{r} K(x) dx
\end{array}
\label{eqsis2}
\end{equation}
Equations \ref{eqsis1} and \ref{eqsis2} form a systems of two equations with two unknowns. The solution of this system provides the $\sigma_{r}^{2}$ and $\beta$ profiles. 

As \cite{biviano2004} pointed out, the practical application of the method needs extrapolations  of the observed quantities $M(r)$, $\Sigma(R)$ and $\sigma^{2}(R)$ to large radii. In addition, Eq. \ref{eqsis2} contains two integrals with the lower integration limit $r=0$. This lower limit requires extrapolation of the functions $r^{3} \Psi(r)$ and $K(r)$ from the innermost measured point to $r=0$ \citep[see also][]{katgert2004}.  In our case, we extrapolated the same fitted analytic profiles of  $M(r)$, $\Sigma(R)$ and $\sigma_{p}(R)$ to $r = 0$ and to large radii. We considered $3 \times r_{200}$ as the upper limit of the integrals. This value is sufficiently far from the last measured point and no differences were observed in the results with farther extrapolations.

Our Jeans Equation Solver (JES) code was tested using analytic solutions (see Appendix A) and data  from cosmological simulations (see Appendix B). We have also analysed how the uncertainties we estimated on $\beta$  depend on the number of galaxies used  (see Appendix C) and the velocity resolution (Appendix D).

The uncertainties on the $\beta$ parameter have been obtained by adding in quadrature the uncertainties on the mass models and Poisson noise (see Appendix C). The asphericity of the clusters can also affect the values of $\beta$. However, no uncertainties deriving from the lack of sphericity were considered here (see Appendix B for a discussion). 

%
 
 \section{Results}
 
In this section, we present the results on the velocity anisotropy profile of  A\,85. We have analysed the orbital anisotropy of the full population of galaxies and of some subsamples selected according to their luminosity and $g-r$ colour.
 
 \subsection{Radial anisotropy profile of the  galaxy population in A\,85}
 
 Figure \ref{aniso_all} shows the radial anisotropy profile computed with the 358 cluster members not located in substructures. The profile  was obtained using the three different mass models reported in Sec. 2.5 and shown in  Tab. \ref{tabnfw}.  The shaded areas show the uncertainty on the $\beta$ parameter. The solid lines represent the values of $\beta$ for the mean mass model.
 
 
 The anisotropy profile of the full galaxy population is an increasing function of radius. Galaxies located in the central regions of the cluster ($r/r_{200} < 0.3$) show $\beta \sim 0$ indicating an isotropic velocity distribution. In contrast, galaxies are on increasingly radial orbits in the cluster outer regions. The overall shape of $\beta(r)$  is typical of systems formed by gravitational collapse \citep[see][]{vanalbada1982}. Similar profiles have also been obtained in individual nearby or intermediate-redshift clusters \citep[see][]{biviano2013,annunziatella2014} and in cosmological simulations  \citep[see][]{lemze2012,iannuzzi2012}. The fact that there are no galaxies on radial orbits in the central region of the cluster may indicate that galaxies on these orbits either merged with the central galaxy or were destroyed by strong gravitational interactions with the cluster potential, or they spent enough time in the cluster to have their orbits circularized. 
  
\cite{wojtak2010} obtained the anisotropy profile of galaxies in a sample of 41 nearby galaxy clusters. They concluded that galaxy orbits are isotropic in the cluster centre and radially anisotropic at the cluster virial radius, with the level of anisotropy depending on the evolutionary state of the cluster. These results are in agreement with the anisotropy profile we obtained. In Fig. \ref{aniso_all}, we have overplotted the two values of $\beta$ determined by \cite{wojtak2010} for A\,85 in the center and at $r_{200}$.

\subsection{Velocity anisotropy profiles of red and blue galaxies}

\begin{figure}
   \centering
\includegraphics[width=\hsize]{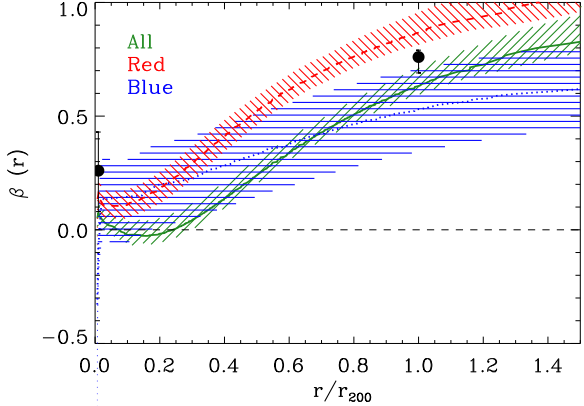}
   \caption{Anisotropy radial profile of all (green), red (red dash) and blue (blue dot) galaxies not located in the substructures of A\,85. The shaded areas correspond to the uncertainties in the values of $\beta$. The solid lines represent $\beta$ obtained from the mean mass model of the cluster (see the text for more details).}
             \label{aniso_all}%
    \end{figure}

The blue and red shaded areas shown in Fig. \ref{aniso_all} are the anisotropy profile for the blue (102 members) and red (256 members) galaxies of A\,85, respectively. The galaxies were classified according to their $g-r$ colour \citep[see ][for the selection method]{agulli2016}.   The uncertainties on the $\beta$ parameter of the blue galaxy population are larger than those of the red population due to the smaller number of galaxies in the sample. However, the mean mass model for both families of galaxies show $\beta \sim 0.1$ in the central cluster region ($r/r_{200} < 0.3$) and larger radial anisotropies at larger radii. In addition, the blue population shows smaller anisotropy values at large clustercentric distances than the red population.

\subsection{$\beta(r)$ for galaxies with different luminosities}

\begin{figure}
   \centering
\includegraphics[width=\hsize]{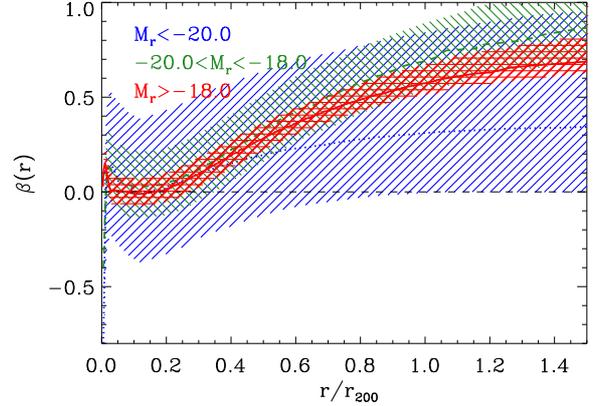}
   \caption{Anisotropy radial profile of galaxies selected by different luminosity cuts and not located in substructures. The shaded areas are similar to Fig. \ref{aniso_all}. The dot blue, dash green and full red lines correspond to galaxies with $M_{r} < -20.0$, $-20.0 < M_{r} < -18.0$ and $M_{r} > -18.0$, respectively.}
             \label{aniso_lum}%
    \end{figure}

We analysed the velocity anisotropy for different luminosity subsets of the A\,85 members. Three luminosity bins were considered. The first group of galaxies is formed by those  with $M_{r} < -20.0$ (we will call them the bright population). The second group are those with  $-20.0 < M_{r} < -18.0$ (intermediate population). The last group is formed by the dwarf galaxy population ($M_{r} > -18.0$). Figure \ref{aniso_lum} shows the radial anisotropy profiles for these three families of galaxies. The $\beta$ parameter of the mean mass model shows that  in the central region of the cluster ($r/r_{200} < 0.3$) the three families show $\beta \sim 0$, which indicates isotropic orbits. At larger radii the anisotropy parameter of the mean mass models shows that bright galaxies have  smaller values of $\beta$ than the other two populations. Intermediate and dwarf galaxies have similar $\beta(r)$, with the intermediate galaxies having slightly larger values of anisotropy. However, the differences in the $\beta$ parameter of the three galaxy samples are not statistically significant. This result is a consequence of the large uncertainties shown by the bright galaxy population, mostly due to the small size of the sample.


The dwarf galaxy population in A\,85 shows an increasingly anisotropy towards the external regions of the cluster. Dwarf galaxies at large radii are located on more radial orbits.   \cite{adami2009} reported the orbital structure of a sample of $\sim 70$ dwarfs located in the central regions ($r/r_{200} < 0.3$) of the Coma cluster. Their orbital analysis rejects the hypothesis of isotropy of these dwarfs in the central region of the cluster. This result is different from what we obtain for A\,85. Our results are in better agreement with those obtained by \cite{annunziatella2016} for an intermediate-redshift cluster. They found that the dwarf population of galaxies of this cluster shows an increasingly $\beta(r)$ profile at large radii. In addition, in the central cluster region, dwarf galaxies were located on tangential orbits.  The absence of dwarf galaxies on radial orbits in the central regions of A\,85 could be related to a selective disruption of low-mass halos within the cluster potential: dwarf galaxies located on radial orbits would be strongly perturbed when they pass through the cluster central region and might not be able to survive \citep[see][]{smith2015}. It could also be possible that the dwarfs we see today in the inner cluster region were accreted by the cluster at early times when the cluster had only assembled a fraction of its present mass. The small value of $\beta$ shown by these galaxies could be a consequence of the circularization of their orbits over time.

     \begin{figure}
   \centering
\includegraphics[width=\hsize]{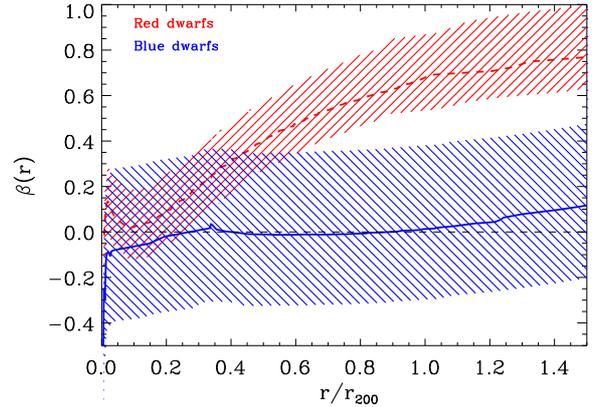}
   \caption{Anisotropy radial profile of blue and red dwarf galaxies of A\,85. The shaded areas are similar to Fig. \ref{aniso_all}. The dash red and full blue lines correspond to red dwarf and blue dwarf galaxies, respectively}
             \label{aniso_dwarf}%
    \end{figure}

\subsection{$\beta(r)$ for red and blue dwarf galaxies}

The large number of dwarf galaxies (189) with velocity information in A\,85 enables us to obtain the radial anisotropy profile for the 117 red and 72 blue dwarf galaxies. Figure \ref{aniso_dwarf} shows $\beta(r)$ for the two families of galaxies. The $\beta$ profile from the mean mass models shows that, in the inner cluster region, blue and red dwarf galaxies show $\beta \sim 0$. In contrast,  the red dwarf galaxies show more radial orbits in the external cluster region. The blue dwarf galaxies show smaller $\beta$ than the red ones at all radii. The correlation between $\beta$ and colour at large clustercentric radius for dwarf galaxies is significant despite the large uncertainties.

The observed strong correlation between the velocity anisotropy and the colour of the dwarf galaxies indicates that the physical processes modifying the stellar colour of these galaxies strongly depends on the orbit of the galaxies: the observed blue dwarf galaxies could be located on orbits which imply smaller interaction with the cluster potential than red dwarfs. This difference can occur if blue dwarf galaxies are located on orbits with large pericenters, so that they never pass near the cluster centre. In contrast, red dwarfs would be located on radial orbits with small pericenters. These galaxies would spend some time close to the cluster centre where the effect of the environment is more pronounced. This difference would indicate that the colour transformation of the dwarf galaxies would be produced near or within the cluster central region.

\subsection{$Q(r)$, $Q_{r}(r)$, and the $\beta - \gamma$ relation}

Numerical simulations of virialized DM halos show  a universal relation between their mass density, $\rho$, and velocity dispersion, $\sigma$, profiles. Thus, the quantity $Q(r) = \rho(r)/\sigma^{3}(r)$ can be fitted by a power-law $r^{-\alpha}$ with $\alpha \sim 1.9 \pm 0.05$ \citep[see][]{taylor2001,rasia2004,ascasibar2004,dehnen2005,vass2009,tissera2010}. This universal $\rho - \sigma$ relation is valid for halos spanning a large range of masses (from galaxy to cluster-size masses), dynamical states, and over their whole resolved radial range.  This relation is also established for the similar quantity $Q_{r}(r) = \rho/\sigma_{r}^{3}(r) \sim r^{-\alpha}$ with $\alpha = 1.94$, being $\sigma_r$ the radial component of the velocity dispersion \citep[see][]{dehnen2005}. Relations similar to these were predicted by the self-similar solution of the spherical infall model of \cite{bertschinger1985}. 

Although $Q(r)$ and $Q_{r}(r)$ have dimensions of phase-space density, they are not related to the coarse-grain phase-space density. They rather 
reflect the distribution of entropy of the DM halos \citep[see][]{vass2009}.   Little is known about these relations from observations. \cite{biviano2013} reported $Q(r)$ and $Q_{r}(r)$ for an intermediate-redshift cluster. They found that the power-law behaviours of $Q(r)$ and $Q_{r}(r)$ were in general reproduced for this cluster as the simulations predicted. Only the star forming galaxies of the cluster did not follow the theoretical relations of $Q(r)$ and $Q_{r}(r)$. They argued that while passive galaxies underwent violent relaxation and have reached dynamical equilibrium, star forming galaxies did not \citep[see][]{biviano2013}. The radial dependence of $Q(r)$ and $Q_{r}(r)$ was also studied  for the nearby cluster A\,2142 \citep[see][]{munari2014}. They found that the $Q(r)$ and $Q_{r}(r)$ radial profiles of all the galaxies of A\,2142 were marginally consistent with the theoretical profiles. In contrast, the agreement improved when only red galaxies were considered. They argued that the differences are due to the blue galaxy population which are recent infall cluster members and have had no time to reach equilibrium within the cluster potential. 

\begin{figure*}
   \centering
\includegraphics[width=\hsize]{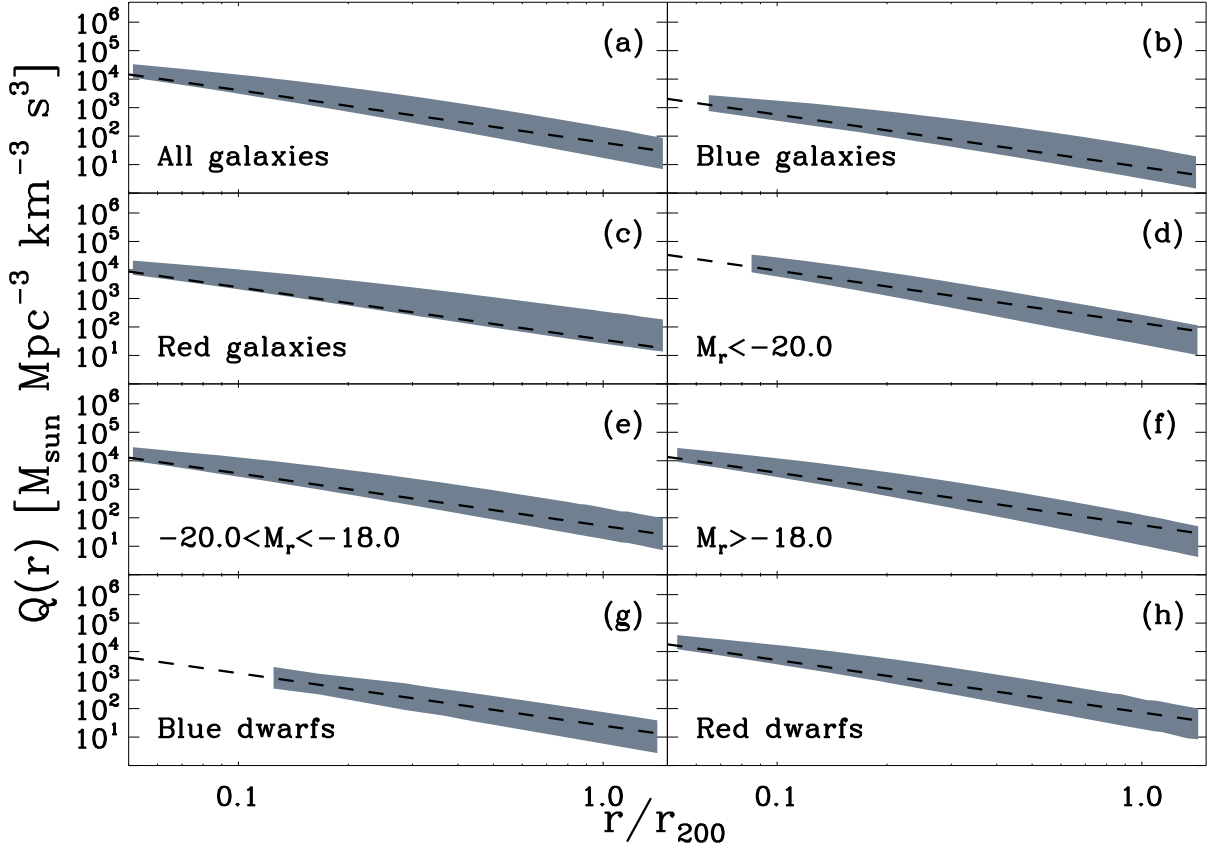}
   \caption{Pseudo phase-space density profiles $Q(r) = \rho/\sigma^{3}$ for all the galaxies and those selected by their $g - r$ colour and luminosities. The dashed lines are the best-fit relations with fixed slope $Q(r) \propto r^{-1.84}$ from \citet{dehnen2005}. The shaded areas show the maximum and minimum values of $Q(r)$ obtained at each radius from the different mass models. }
             \label{q}%
    \end{figure*}

\begin{figure*}
   \centering
\includegraphics[width=\hsize]{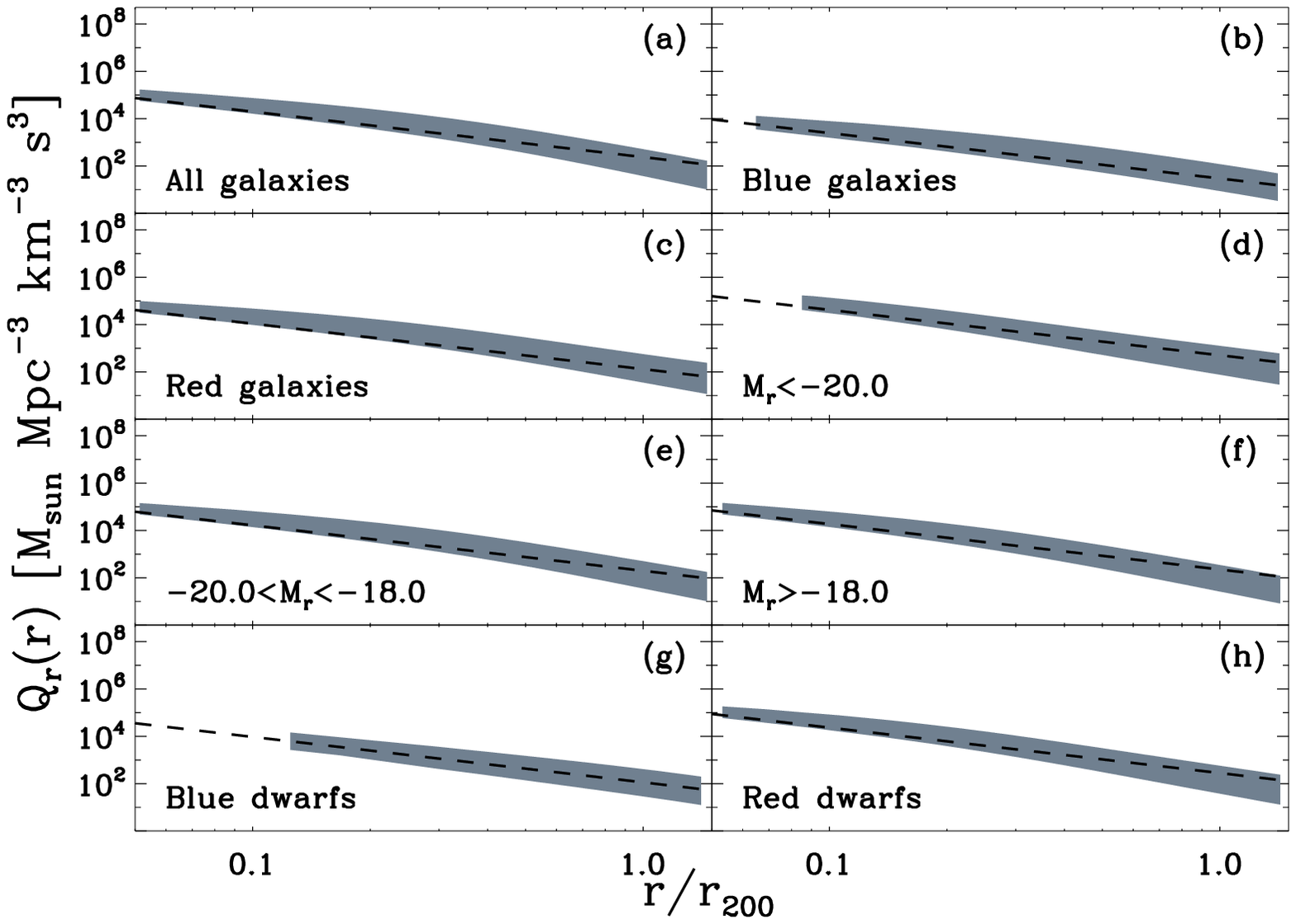}
   \caption{Pseudo phase-space density profiles $Q_{r}(r) = \rho/\sigma_{r}^{3}$ for all the galaxies and those selected by their $g - r$ colour and luminosities. The dashed lines are the best-fit relations with fixed slope $Q_{r}(r) \propto r^{-1.92}$ from \citet{dehnen2005}. The shaded areas show the maximum and minimum values of $Q_{r}(r)$ obtained at each radius from the different mass models.}
             \label{qr}%
    \end{figure*}

Here, we have the information required to study these quantities for A\,85. Figures \ref{q} and \ref{qr} show the values of $Q(r)$ and $Q_{r}(r)$ for different families of galaxies in A\,85. The uncertainties shown in Figs. \ref{q} and \ref{qr} for $Q(r)$ and $Q_{r}(r)$ take into account the different mass models of the cluster. Including Poisson noise in the estimate of the uncertainty leaves the agreement with the Dehnen $\&$ McLaughlin best fit unaffected. In all cases, the observed $Q(r)$ and $Q_{r}(r)$ follow the theoretical power law relations obtained in the simulations. We can conclude that the galaxies of A85, independently of their luminosity or stellar colour, have reached dynamical equilibrium within the cluster potential. 

\cite{hansen2006} found a correlation between the DM radial density slope, $\gamma(r)$, and the velocity anisotropy for structures in dynamical equilibrium. In addition, the universality of this relation could indicate that the mass density and anisotropy profiles are also universal. Later works established that the relation between $\beta$ and $\gamma$ was only in the inner part of the halos ($r<0.3 r_{vir}$). No theoretical correlation was found for larger radii \citep[see][]{tissera2010, lemze2012}. The only constraint found at all radii between $\beta$ and $\gamma$  was $\gamma < \beta$  \citep[see][]{an2006,ciotti2010}.

From the observational point of view, \cite{biviano2013} found that the theoretical relation between $\beta$ and $\gamma$ was recovered for the galaxies of the intermediate-redshift cluster MACS J1206.2-0847.  \cite{munari2014} found instead that the $\beta - \gamma$ relation for the red galaxy population of the nearby cluster A\,2142 matches the theoretical relation only in the inner cluster region.

\begin{figure}
   \centering
\includegraphics[width=\hsize]{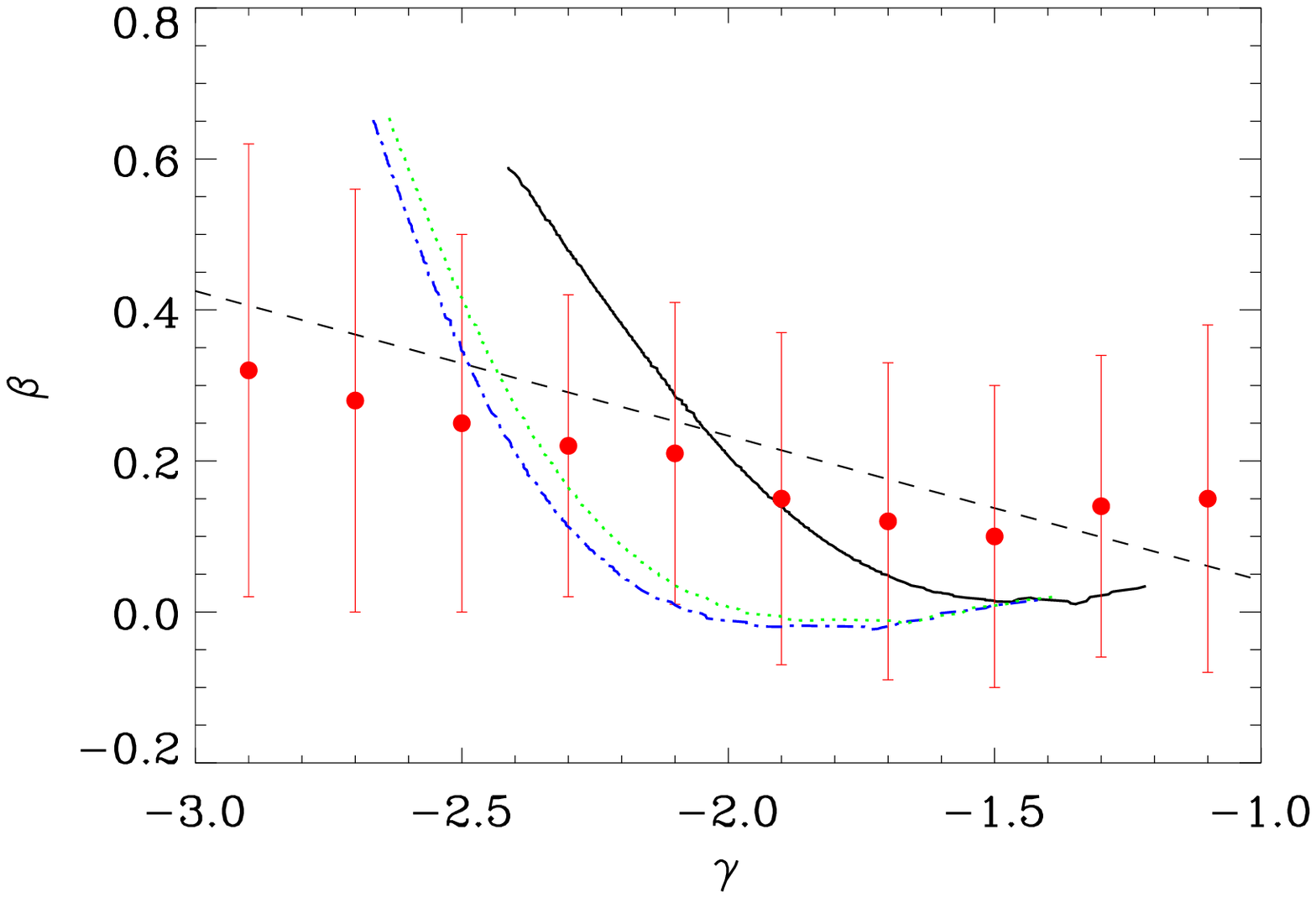}
   \caption{$\beta - \gamma$ relation for all the galaxies in A\,85 for the different mass models considered in this work. The blue dash-dot, green dot and black lines represent the $\gamma - \beta$ relation obtained from the \citet{durret2005},  \citet{rines2006}, and \citet{sandersen2003} mass models, respectively. The black dashed line is the relation obtained by \citet{hansen2006}. The red points show the relation obtained by \citet{lemze2012}.}
             \label{beta_gamma}%
    \end{figure}

Figure \ref{beta_gamma} shows the $\beta - \gamma$ relation for the galaxies of A\,85. We also overplotted the linear $\beta - \gamma$ relation obtained by \cite{hansen2006} and the one derived from  numerical simulations by \cite{lemze2012}. In the case of A\,85 there is a poor match between our measure and the expected  $\beta - \gamma$ relation. This result could indicate that this relation is not universal and that the  two quantities are weakly correlated, as pointed out by \cite{tissera2010} and \cite{lemze2012}. Nevertheless, $\gamma < \beta$ is confirmed for A\,85.

 \section{Discussion}
 
 The present work analysed  the velocity anisotropy profiles for the galaxies in the A\,85 cluster. The identification of a large number of cluster members enabled us to estimate $\beta$ for different samples of galaxies, including the dwarf population. Here, we  discuss our results. In particular, we discuss the dependence of $\beta$ on the galaxy substructure, luminosity and stellar colour. Taking into account the results obtained by \cite{agulli2014} and \cite{agulli2016} on this cluster, we give a global picture of the galaxy evolution in A\,85.
 
 \subsection{Variation of $\beta$ and galaxy substructure}
 
 One of the assumptions of Jeans equation is the virialization of the system. The cluster A\,85 is not fully virialized. Several imprints indicate the presence of substructures in this system \citep[see e.g.,][]{ramella2007,aguerri2010,agulli2016,yu2016}. We presented the results only considering the galaxies not in substructures. However, we wonder how the results on $\beta$ change by including the galaxies in substructures: Are the galaxies in substructures on more radial or tangential orbits than the relaxed population? To answer this question, we analysed the velocity anisotropy profile obtained by taking into account the unrelaxed galaxy population.
 
\begin{figure}
   \centering
\includegraphics[width=\hsize]{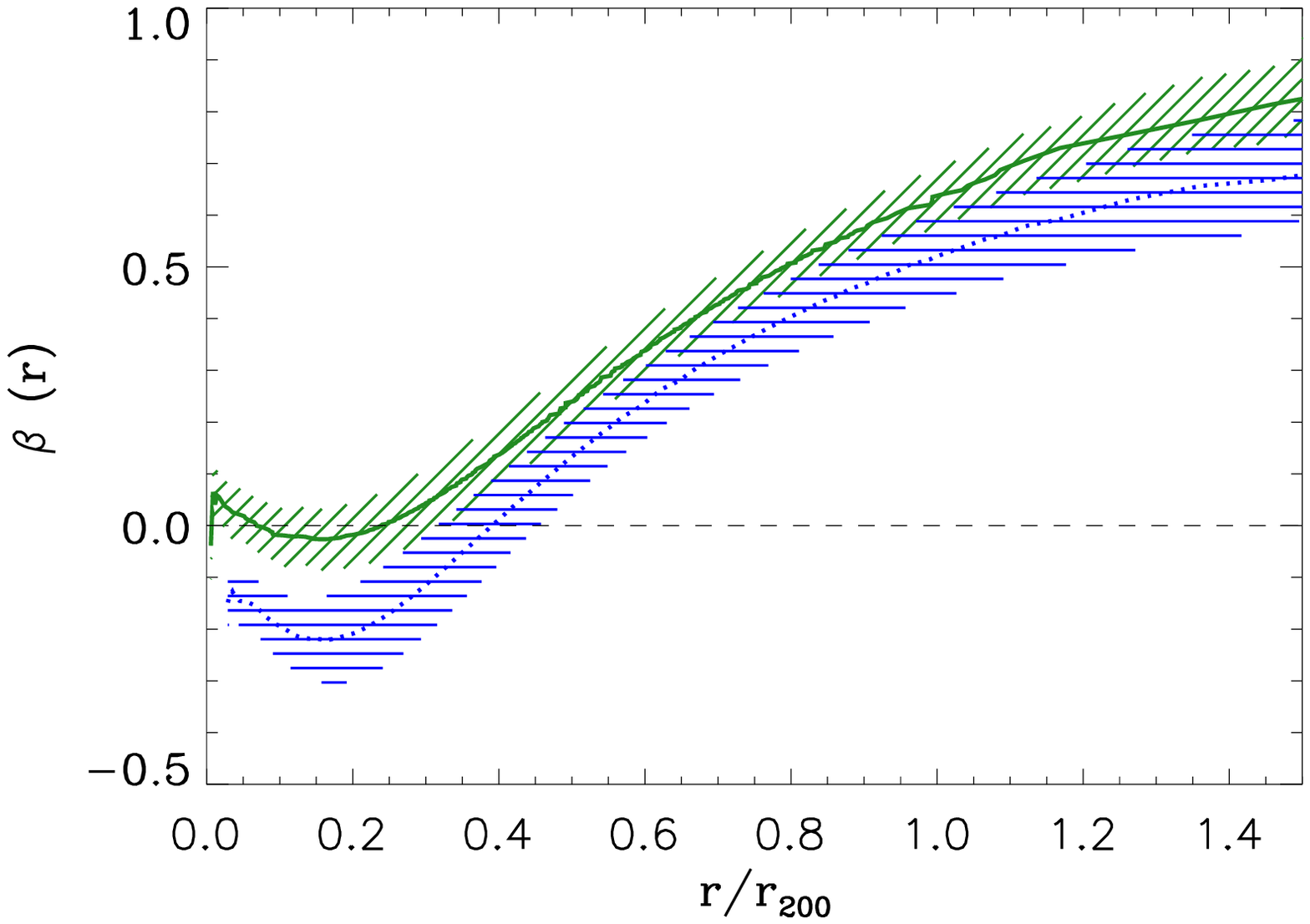}
   \caption{Velocity anisotropy profile of all the galaxies of A\,85, including galaxies in substructures (blue dot line), and velocity anisotropy profile of all the galaxies not in substructures (green full line). The shaded areas and solid lines are as in Fig. \ref{aniso_all}.}
             \label{sub_nosub}%
    \end{figure}

 Figure \ref{sub_nosub} shows a comparison of $\beta$ radial profiles obtained by taking into account either all the cluster members or the virialized galaxies alone. The shapes of the two anisotropy profiles are similar.  The main difference between the two profiles is that $\beta$ is smaller when the galaxies in substructure are included in the sample. This difference indicates that galaxies in substructure have less radial orbits than the virialized galaxy population.  
 
 These results are in agreement with other observational results. \cite{biviano2004} found that galaxies in substructures show a tangential orbital distribution. Similar results were obtained by \cite{hwang2008}. These results are also similar to those obtained with numerical simulations. In particular, the study by \cite{lemze2012} on the velocity anisotropy profiles of simulated DM halos shows that unrelaxed systems have lower values of $\beta$. They interpreted these results as an indication that DM particles are accreted on less radial orbits than those in relaxed halos. In addition, \cite{taylor2004} studied the evolution of the orbits of infalling substructures. They concluded that substructures falling on more radial orbits are disrupted sooner than those on tangential ones. 
  
 Our findings indicate that in clusters like A\,85 with a small percentage of galaxies in substructures, the inclusion in the sample of the non-virialized galaxies slightly affects the values of $\beta$ but not its dependence on radius. 
 
\subsection{Variation of $\beta(r)$ with the galaxy properties}
 


The other important correlation observed in the cluster is between the orbital anisotropy and the stellar colour of the galaxies. We found that blue galaxies are on less radial orbits than red galaxies. This observation holds for both the full sample and for the dwarf galaxies alone. The former sample shows $\beta > 0$ at all radii, being $\beta \sim 0.5$ at $r=r_{200}$. In the case of the blue dwarf galaxies, $\beta \sim 0$ at all radii. These results suggests that bright blue galaxies are on more radial orbits than blue dwarfs.

Previous studies in the literature reported different results. Studies using stacked clusters found that  early-type and late-type spiral galaxies show different orbital structures, with the latter on more radially-elongated orbits at large radii \citep[see][]{biviano2004}. Assuming that late-type galaxies are basically blue, this result disagrees with the result obtained for A\,85. In contrast, results based on individual clusters show that red and blue galaxies have similar anisotropy profiles \citep[see][]{hwang2008}. These differences together with our results could indicate that the orbital structure of the red and blue galaxies in galaxy clusters is not universal.

Our findings agree with theoretical results on the orbital structure of red and blue galaxies. In particular,  \cite{iannuzzi2012} analysed the anisotropy evolution of red and blue galaxies in clusters  using data from the Millennium Simulation. They found that  blue galaxies show significantly lower anisotropy than red ones at all redshifts. In addition, the galaxies that are blue at $z=0$ are falling into the cluster with the most tangential orbits \citep[see][]{iannuzzi2012}. 

In the literature, it has been argued that blue galaxies were recently accreted onto clusters and did not have enough  time to reach virialization and change their stellar colour. This argument was based on the observed larger velocity dispersion of the blue galaxy population in clusters \citep[see e.g.][]{aguerri2007} or by the different shape of the radial velocity dispersion profiles of blue and red galaxies \citep[see e.g.][]{adami1998}. However, \cite{goto2005} argued against this hypothesis. 

Our results for A\,85 indicate that blue members reached a virialization state. Our conclusion is based on the behaviour of $Q(r)$ and $Q_{r}(r)$. Blue galaxies of A\,85 show values of $Q(r)$ and $Q_{r}(r)$ similar to those obtained from the theoretical relations inferred for virialized DM halos. Therefore, blue galaxies of A\,85 had enough time to changed their stellar colours and we may rule out the argument based on the recent accretion. The reason why they still are blue could be due to their orbital distribution within the cluster.

Galaxies on radial orbits could have small pericenters and pass closer to the cluster centre than galaxies on more tangential orbits. Galaxies passing close to the cluster centre could suffer strong transformation due to tidal interactions and gas stripping. These mechanisms could produce large loss of stellar mass \citep[see][]{aguerri2009,smith2015} or gas mass \citep[see][]{quilis2000}. This kind of interactions can transform galaxies, by quenching their star formation and reddening their stellar populations. Thus, we can expect that  blue galaxies with infall radial orbits switched off their star formation and were transformed into red and passive galaxies. The blue galaxies observed today in the cluster are  those on less radial orbits and are thus less affected by the dense environment of the cluster centre. Galaxies on these orbits can keep their gas reservoirs and maintain their star formation active for longer time. The orbital - colour correlation observed today in A\,85 would be a consequence of a selected transformation of galaxies due to their orbital properties.

\cite{biviano2013} analysed the orbital structure of galaxies in the intermediate-redshift cluster MACS J1206.2-0847 ($z = 0.44$). This massive cluster ($\sigma  = 1087$ km s$^{-1}$) shows no differences in $\beta$ for its blue and red galaxies. Both, galaxy populations are located on increasingly radial orbits towards larger cluster radii. The orbital differences observed between A\,85 and this intermediate-redshift cluster could be interpreted as consequence of evolution: blue galaxies on radial orbits at $z=0.44$  lost their gas, switched off their star formation and reddened their stellar colour. Clearly, a larger number of studies of the orbital properties of clusters at intermediate and low redshifts are needed in order to fully support this statement.

 \subsection{A global picture of A\,85}
 
 This paper is the third in a series on the properties of the galaxy population of A\,85. In the previous two papers we have analysed the LF of the cluster \citep[see][]{agulli2014, agulli2016}. In particular, \cite{agulli2014} found that the global LF of A\,85 has an upturn at faint magnitudes showing a faint slope similar to the LF of field galaxies. The main difference is that the red dwarf galaxies dominate the faint-end of the LF in the cluster while blue dwarfs dominate in the field. This result was interpreted as a consequence of  the quenching of the star formation in dwarf galaxies by the cluster environment. The different orbital anisotropy found here for  blue and red dwarfs agree with this interpretation. Red dwarfs are on more radial orbits than blue ones. They passed near the cluster centre on eccentric orbits with small pericenter and suffered gravitational and hydrodynamical interactions. In contrast, the blue dwarf galaxies observed today are on more circular orbits and only mildly suffered the interaction with the environment. Our observational result agrees with the velocity anisotropy profiles in numerical simulations \citep[see][]{iannuzzi2012}. We can conclude that the reddening of the dwarf galaxies in A\,85 is a consequence of the different orbits they have. In other words, galaxies falling into the cluster on radial orbits are reddened. In contrast, those falling on more circular orbits would keep their stellar colours unchanged for longer time and would be the galaxies observed as blue galaxies today.

 
 
\section{Conclusions}

We have analysed the orbital structure of the galaxies in the massive and nearby cluster A\,85. The main results obtained are:

\begin{itemize}
\item The galaxy population of the cluster preferentially has isotropic orbits in the innermost region ($r/r_{200} < 0.3$). Outside this region galaxies are on increasingly radial orbits at larger and larger radii.

\item Blue and red galaxies are on orbits with different shape: blue galaxies are on less radial orbits than red galaxies. This result is independent of the galaxy luminosity, except for blue dwarf galaxies that have $\beta \sim 0$ at all radii.

\item The different galaxy populations are in equilibrium within the cluster potential. This conclusion is based on our estimate of the space-densities $Q(r)$ and $Q_{r}(r)$ that follow the radial trend of virialized DM halos in $N$-body simulations.

\item Blue galaxies are not recent arrivals to the cluster potential: they have blue colours because their  orbits do not bring them as close to the cluster centre as the more radial orbits of red galaxies. 

\end{itemize}

The results presented in this work indicate that the orbital structure of  galaxies plays a key role in the colour transformation of the cluster members. In the future, we will apply this kind of study to a larger sample of galaxy clusters with different dynamical states and physical properties.

\section*{acknowledgements}
      JALA and IA want to thank Spanish Ministerio de Economia y competitividad (MINECO) under the grant AYA2013-43188-P for supporting this work. CDV acknowledges support from the Ministry of Economy and Competitiveness (MINECO) through grants AYA2013-46886 and AYA2014-58308. IA and AD acknowledge partial support from the INFN grant InDark and the grant PRIN 2012 "Fisica Astroparticellare Teorica" of the Italian Ministry of University and Research. CDV acknowledges financial support from MINECO under the Severo Ochoa Programs SEV-2011-0187 and SEV-2015-0548. Numerical simulations were performed on the Teide High-Performance Computing (Teide-HPC) facilities provided by the Instituto Tecnol\'ogico y de Energ\'{\i}as Renovables (ITER, SA, \url{http://teidehpc.iter.es}). This research has made use of the Sixth Data Release of SDSS, and of the NASA/IPAC Extragalactic Database which is operated by the Jet Propulsion Laboratory, California Institute of Technology, under contract with the National Aeronautics and Space Administration. The WHT and its service programme are operated on the island of La Palma by the Isaac Newton Group in the Spanish Observatorio del Roque de los Muchachos of the Instituto de Astrof\'isica de Canarias.





\begin{thebibliography}{99}
\bibitem[Adami et al.(2009)]{adami2009} Adami, C., Le Brun, V., Biviano, A., et al.\ 2009, \aap, 507, 1225
\bibitem[Adami et al.(1998)]{adami1998} Adami, C., Biviano, A., \& Mazure, A.\ 1998, \aap, 331, 439 
\bibitem[An \& Evans(2006)]{an2006} An, J.~H., \& Evans, N.~W.\ 2006, \apj, 642, 752 
\bibitem[Aguerri et al.(2001)]{aguerri2001} Aguerri, J.~A.~L., Hunter, J.~H., Prieto, M., et al.\ 2001, \aap, 373, 786 
\bibitem[Aguerri et al.(2007)]{aguerri2007} Aguerri, J.~A.~L., S{\'a}nchez-Janssen, R., \& Mu{\~n}oz-Tu{\~n}{\'o}n, C.\ 2007, \aap, 471, 17 
\bibitem[Aguerri \& S{\'a}nchez-Janssen(2010)]{aguerri2010} Aguerri, J.~A.~L., \& S{\'a}nchez-Janssen, R.\ 2010, \aap, 521, A28 
\bibitem[Aguerri \& Gonz{\'a}lez-Garc{\'{\i}}a(2009)]{aguerri2009} Aguerri, J.~A.~L., \& Gonz{\'a}lez-Garc{\'{\i}}a, A.~C.\ 2009, \aap, 494, 891
\bibitem[Agulli et al.(2014)]{agulli2014} Agulli, I., Aguerri, J.~A.~L., S{\'a}nchez-Janssen, R., et al.\ 2014, \mnras, 444, L34 
\bibitem[Agulli et al.(2016)]{agulli2016} Agulli, I., Aguerri, J.~A.~L., S{\'a}nchez-Janssen, R., et al.\ 2016, \mnras,  
\bibitem[Annunziatella et al.(2014)]{annunziatella2014} Annunziatella, M., Biviano, A., Mercurio, A., et al.\ 2014, \aap, 571, A80 
\bibitem[\protect\citeauthoryear{Annunziatella et al.}{2016}]{annunziatella2016} Annunziatella M., et al., 2016, A\&A, 585, A160 
\bibitem[Ascasibar et al.(2004)]{ascasibar2004} Ascasibar, Y., Yepes, G., Gottl{\"o}ber, S., M{\"u}ller, V.\ 2004, \mnras, 352, 1109 
\bibitem[Atrio-Barandela et al.(2008)]{atrio2008} Atrio-Barandela, F., Kashlinsky, A., Kocevski, D., \& Ebeling, H.\ 2008, \apjl, 675, L57 


\bibitem[Bertschinger(1985)]{bertschinger1985} Bertschinger, E.\ 1985, \apjs, 58, 39 
\bibitem[Bialas et al.(2015)]{bialas2015} Bialas, D., Lisker, T., Olczak, C., Spurzem, R., \& Kotulla, R.\ 2015, \aap, 576, A103 
\bibitem[Binney \& Mamon(1982)]{binney1982} Binney, J., \& Mamon, G.~A.\ 1982, \mnras, 200, 361 
\bibitem[Biviano et al.(2013)]{biviano2013} Biviano, A., Rosati, P., Balestra, I., et al.\ 2013, \aap, 558, A1 
\bibitem[Biviano \& Katgert(2004)]{biviano2004} Biviano, A., \& Katgert, P.\ 2004, \aap, 424, 779
\bibitem[Blanton \& Moustakas(2009)]{blanton2009} Blanton, M.~R., \& Moustakas, J.\ 2009, \araa, 47, 159 
\bibitem[Bou{\'e} et al.(2008)]{boue2008} Bou{\'e}, G., Durret, F., Adami, C., et al.\ 2008, \aap, 489, 11
\bibitem[Bravo-Alfaro et al.(2009)]{bravoalfaro2009} Bravo-Alfaro, H., Caretta, C.~A., Lobo, C., Durret, F., \& Scott, T.\ 2009, \aap, 495, 379

\bibitem[Carlberg et al.(1997)]{carlberg1997} Carlberg, R.~G., Yee, 
H.~K.~C., Ellingson, E., et al.\ 1997, \apjl, 485, L13
\bibitem[Ciotti \& Morganti(2010)]{ciotti2010} Ciotti, L., \& Morganti, L.\ 2010, \mnras, 408, 1070 

\bibitem[Davis et al.(1985)]{1985ApJ...292..371D} Davis, M., Efstathiou, G., Frenk, C.~S., \& White, S.~D.~M.\ 1985, \apj, 292, 371
\bibitem[Dehnen \& McLaughlin(2005)]{dehnen2005} Dehnen, W., \& McLaughlin, D.~E.\ 2005, \mnras, 363, 1057 
\bibitem[Demarco et al.(2003)]{demarco2003} Demarco, R., Magnard, F., Durret, F., \& M{\'a}rquez, I.\ 2003, \aap, 407, 437 
\bibitem[Diaferio et al.(2005)]{diaferio2005} Diaferio, A., Geller, M.~J., \& Rines, K.~J.\ 2005, \apjl, 628, L97 
\bibitem[Diaferio(1999)]{diaferio1999} Diaferio, A.\ 1999, \mnras, 309, 610 
\bibitem[Diaferio \& Geller(1997)]{diaferio1997} Diaferio, A., \& Geller, M.~J.\ 1997, \apj, 481, 633 
\bibitem[Durret et al.(2005)]{durret2005} Durret, F., Lima Neto, G.~B., \& Forman, W.\ 2005, \aap, 432, 809 
\bibitem[Durret et al.(2003)]{durret2003} Durret, F., Lima Neto, G.~B., Forman, W., \& Churazov, E.\ 2003, \aap, 403, L29

\bibitem[\protect\citeauthoryear{Fujita}{2004}]{fujita2004} Fujita Y., 2004, PASJ, 56, 29 

\bibitem[Gnedin(2003)]{gnedin2003} Gnedin, O.~Y.\ 2003, \apj, 582, 141
\bibitem[Goto(2005)]{goto2005} Goto, T.\ 2005, \mnras, 359, 1415 
\bibitem[Gunn \& Gott(1972)]{gunn1972} Gunn, J.~E., \& Gott, J.~R., III 1972, \apj, 176, 1

\bibitem[Hahn \& Abel(2011)]{2011MNRAS.415.2101H} Hahn, O., \& Abel, T.\ 2011, \mnras, 415, 2101
\bibitem[Hansen \& Moore(2006)]{hansen2006} Hansen, S.~H., \& Moore, B.\ 2006, \na, 11, 333 
\bibitem[Hwang 
\& Lee(2008)]{hwang2008} Hwang, H.~S., \& Lee, M.~G.\ 2008, \apj, 676, 218-247

\bibitem[Iannuzzi \& Dolag(2012)]{iannuzzi2012} Iannuzzi, F., \& Dolag, K.\ 2012, \mnras, 427, 1024 
\bibitem[\protect\citeauthoryear{Larson, Tinsley, \& Caldwell}{1980}]{larson1980} Larson R.~B., Tinsley B.~M., Caldwell C.~N., 1980, ApJ, 237, 692 

\bibitem[Ichinohe et al.(2015)]{ichinohe2015} Ichinohe, Y., Werner, N., Simionescu, A., et al.\ 2015, \mnras, 448, 2971 

\bibitem[Katgert et al.(2004)]{katgert2004} Katgert, P., Biviano, A., \& Mazure, A.\ 2004, \apj, 600, 657 
\bibitem[Kent \& Gunn(1982)]{kent1982} Kent, S.~M., \& Gunn, J.~E.\ 1982, \aj, 87, 945 
\bibitem[Knebe et al.(2006)]{knebe2006} Knebe, A., Power, C., Gill, S.~P.~D., \& Gibson, B.~K.\ 2006, \mnras, 368, 741 

\bibitem[Lemze et al.(2012)]{lemze2012} Lemze, D., Wagner, R., Rephaeli, Y., et al.\ 2012, \apj, 752, 141 
\bibitem[Lewis et al.(2000)]{2000ApJ...538..473L} Lewis, A., Challinor, A., \& Lasenby, A.\ 2000, \apj, 538, 473

\bibitem[Mamon et al.(2013)]{mamon2013} Mamon, G.~A., Biviano, A., \& Bou{\'e}, G.\ 2013, \mnras, 429, 3079 
\bibitem[M{\'e}ndez-Abreu et al.(2008)]{mendezabreu2008} M{\'e}ndez-Abreu, J., Aguerri, J.~A.~L., Corsini, E.~M., \& Simonneau, E.\ 2008, \aap, 478, 353 
\bibitem[Merrit(1987)]{merritt1987} Merritt, D.\ 1987, \apj, 313, 121 
\bibitem[Moore et al.(1998)]{moore1998} Moore, B., Lake, G., \& Katz, N.\ 1998, \apj, 495, 139 
\bibitem[Mastropietro et al.(2005)]{mastropietro2005} Mastropietro, C., Moore, B., Mayer, L., et al.\ 2005, \mnras, 364, 607
\bibitem[Munari et al.(2014)]{munari2014} Munari, E., Biviano, A., \& Mamon, G.~A.\ 2014, \aap, 566, A68 

\bibitem[Navarro et al.(1997)]{navarro1997} Navarro, J.~F., Frenk, C.~S., \& White, S.~D.~M.\ 1997, \apj, 490, 493 

\bibitem[Quilis et al.(2000)]{quilis2000} Quilis, V., Moore, B., \& Bower, R.\ 2000, Science, 288, 1617

\bibitem[Ramella et al.(2007)]{ramella2007} Ramella, M., Biviano, A., Pisani, A., et al.\ 2007, \aap, 470, 39 
\bibitem[Rasia et al.(2004)]{rasia2004} Rasia, E., Tormen, G., \& Moscardini, L.\ 2004, \mnras, 351, 237 
\bibitem[Reiprich \& B\"{o}hringer(2002)]{reiprich2002} Reiprich, T.~H., \& B\"{o}hringer, H.\ 2002, \apj, 567, 716
\bibitem[Rines et al.(2003)]{rines2003} Rines, K., Geller, M.~J., Kurtz, M.~J., \& Diaferio, A.\ 2003, \aj, 126, 2152 
\bibitem[Rines \& Diaferio(2006)]{rines2006} Rines, K., \& Diaferio, A.\ 2006, \aj, 132, 1275 

\bibitem[S{\'a}nchez et al.(2013)]{2013MNRAS.433.1202S} S{\'a}nchez, A.~G., Kazin, E.~A., Beutler, F., et al.\ 2013, \mnras, 433, 1202 
\bibitem[Sanderson \& Ponman(2003)]{sandersen2003} Sanderson, A.~J.~R., \& Ponman, T.~J.\ 2003, \mnras, 345, 1241 
\bibitem[Serra \& Diaferio(2013)]{serra2013} Serra, A.~L., \& Diaferio, A.\ 2013, \apj, 768, 116 

\bibitem[Serra et al.(2011)]{serra2011} Serra, A.~L., Diaferio, A., Murante, G., \& Borgani, S.\ 2011, \mnras, 412, 800 
\bibitem[Slee et al.(2001)]{slee2001} Slee, O.~B., Roy, A.~L., Murgia, M., Andernach, H., \& Ehle, M.\ 2001, \aj, 122, 1172
\bibitem[Schmidt \& Allen(2007)]{schmidt2007} Schmidt, R.~W., \& Allen, S.~W.\ 2007, \mnras, 379, 209 
\bibitem[Schenck et al.(2014)]{schenck2014} Schenck, A., Park, S., Burrows, D.~N., et al.\ 2014, \apj, 791, 50
\bibitem[Simonneau et al.(1998)]{simonneau1998} Simonneau, E., Varela, A.~M., \& Munoz-Tunon, C.\ 1998, Nuovo Cimento B Serie, 113, 927 
\bibitem[Smith et al.(2015)]{smith2015} Smith, R., 
S{\'a}nchez-Janssen, R., Beasley, M.~A., et al.\ 2015, \mnras, 454, 2502 
\bibitem[Smith et al.(2010)]{smith2010} Smith, R., Davies, J.~I., \& Nelson, A.~H.\ 2010, \mnras, 405, 1723 
\bibitem[Solanes \& Salvador-Sole(1990)]{solanes1990} Solanes, J.~M., \& Salvador-Sole, E.\ 1990, \aap, 234, 93 
\bibitem[Springel et al.(2001)]{2001MNRAS.328..726S} Springel, V., White, S.~D.~M., Tormen, G., \& Kauffmann, G.\ 2001, \mnras, 328, 726 
\bibitem[Springel(2005)]{2005MNRAS.364.1105S} Springel, V.\ 2005, \mnras, 364, 1105 
\bibitem[Stark(1977)]{stark1977} Stark, A.~A.\ 1977, \apj, 213, 368 

\bibitem[Taylor \& Babul(2004)]{taylor2004} Taylor, J.~E., \& Babul, A.\ 2004, \mnras, 348, 811 
\bibitem[Taylor \& Navarro(2001)]{taylor2001} Taylor, J.~E., \& Navarro, J.~F.\ 2001, \apj, 563, 483 
\bibitem[Tissera et al.(2010)]{tissera2010} Tissera, P.~B., White, S.~D.~M., Pedrosa, S., \& Scannapieco, C.\ 2010, \mnras, 406, 922 
\bibitem[Trujillo et al.(2002)]{trujillo2002} Trujillo, I., Asensio Ramos, A., Rubi{\~n}o-Mart{\'{\i}}n, J.~A., et al.\ 2002, \mnras, 333, 510 

\bibitem[Umetsu et al.(2010)]{umetsu2010} Umetsu, K., Medezinski, E., Broadhurst, T., et al.\ 2010, \apj, 714, 1470 

\bibitem[van Albada(1982)]{vanalbada1982} van Albada, T.~S.\ 1982, \mnras, 201, 939 
\bibitem[Vass et al.(2009)]{vass2009} Vass, I.~M., Valluri, M., Kravtsov, A.~V., \& Kazantzidis, S.\ 2009, \mnras, 395, 1225 
\bibitem[Vikhlinin et al.(2006)]{vikhlinin2006} Vikhlinin, A., Kravtsov, A., Forman, W., et al.\ 2006, \apj, 640, 691 


\bibitem[Wojtak \& {\L}okas(2010)]{wojtak2010} Wojtak, R., \& {\L}okas, E.~L.\ 2010, \mnras, 408, 2442 

\bibitem[\protect\citeauthoryear{Yu et al.}{2016}]{yu2016} Yu H., Diaferio A., Agulli I., Aguerri J.~A. L., Tozzi P., 2016, arXiv, arXiv:1609.02237 
\bibitem[Yu et al.(2015)]{yu2015} Yu, H., Serra, A.~L., Diaferio, A., \& Baldi, M.\ 2015, \apj, 810, 37 

\bibitem[Zarattini et al.(2015)]{zarattini2015} Zarattini, S., Aguerri, J.~A.~L., S{\'a}nchez-Janssen, R., et al.\ 2015, \aap, 581, A16 
\end{thebibliography}




\appendix


\section{The anisotropy profile for an analytic case}

We run several tests in order to check JES, the code we developed for the computation of the velocity anisotropy parameter in galaxy clusters. Here, we present the anisotropy parameter starting from an analytic function \citep[see][]{solanes1990}. 

We assume that the galaxy volumetric density is given by an analytical King law :
\begin{equation}
\nu_{g}(r)= \nu_{0} (1 + r^{2})^{-3/2}
\end{equation}
being $\nu_{0}$ the central galaxy density, which is equal to $\Sigma_{0}/2$ in terms of the observed central projected galaxy number density. Using this law for $\nu_{g}$ we can compute the projected galaxy density, $\Sigma(R)$, by the transform
\begin{equation}
\Sigma(R) = 2 \int_{R}^{\infty} \frac{r \nu_{g}(r) dr}{\sqrt{r^{2} -R^{2}}}
\end{equation}
We also assume that the mass profile of the system is 
\begin{equation}
M(r) = M_{c} (ln (r+u) -r/u)
\label {masana}
\end{equation}
where $M_{c}$ is the normalization {\citep[see][]{solanes1990}  and $u(r) = (1 +r^{2})^{1/2}$. Moreover, we assume an analytic expression for $H(R) = H(0) (1 + R^{2})^{-1}$ with $H(0) = \Sigma_{0} \sigma_{p}^{2}(0)/2$. In this case, $\sigma_{p}^{2}(0)$ represents the central projected velocity dispersion of the system. We can infer it by using Eq. \ref{eqh}: $\sigma_{p}^{2}(R)= 2 H(R)/\Sigma(R)$.

Using the above profiles of $\nu_{g}$, $M(r)$ and $H(R)$, we have the velocity anisotropy profile shown in Fig. \ref{betaana}, assuming $\Sigma_{0}= 100$ Gal Mpc$^{-2}$ and $\sigma_{0}$ = 1000 km s$^{-1}$. 
In Fig. \ref{betaana} we also show the anisotropy profile we obtain with our code, with $\Sigma_{0}= 100 $ Gal Mpc$^{-2}$ and $\sigma_\text{los,0}$ = 1000 km s$^{-1}$. A visual inspection confirms the agreement between the two curves.


   \begin{figure}
   \centering
\includegraphics[width=\hsize]{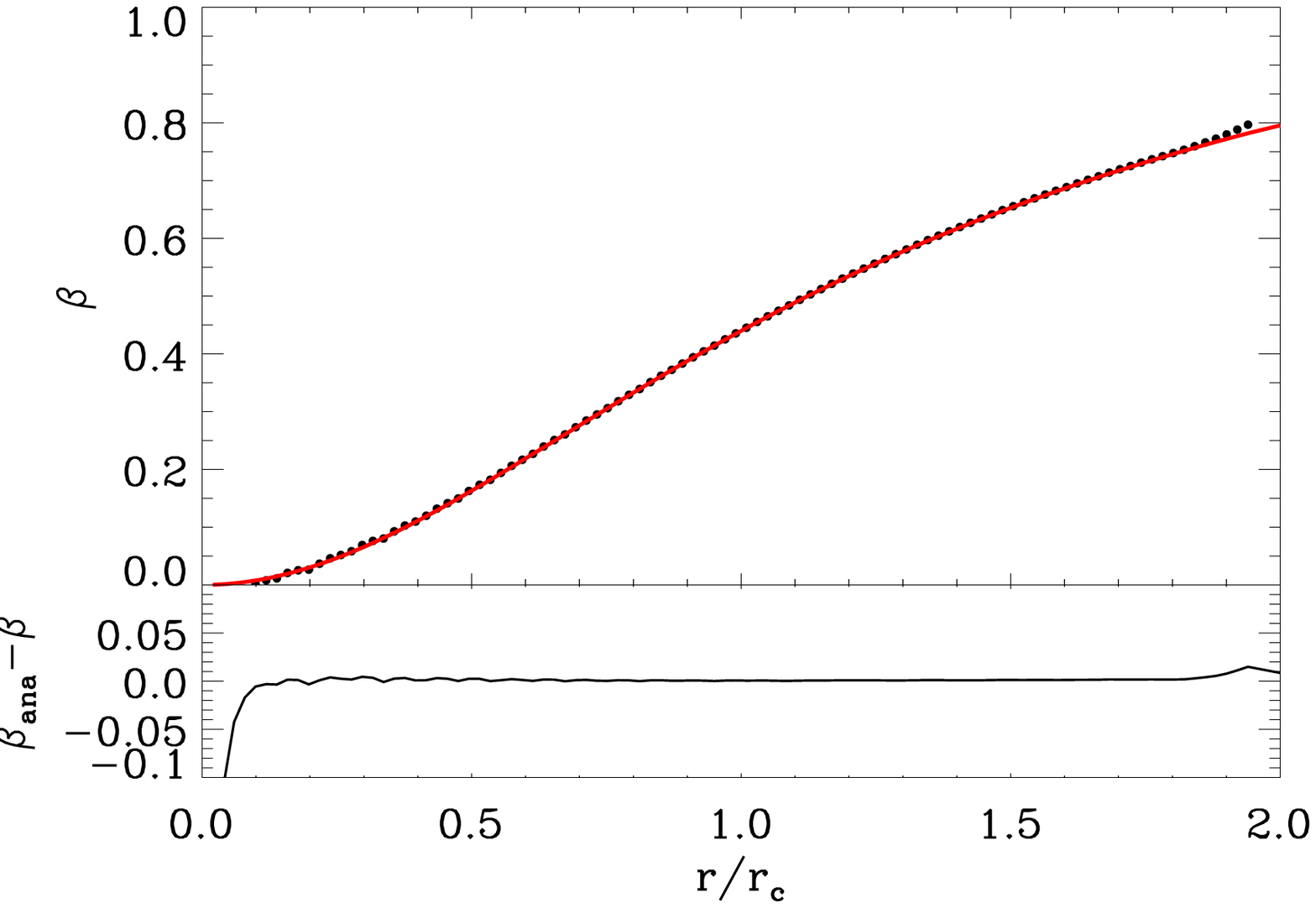}
   \caption{Anisotropy profile obtained from the analytic case (black points). The red line represents the anisotropy profile obtained by using the solution of the Jeans equation by our code.}
             \label{betaana}%
    \end{figure}

\section{The anisotropy profile from galaxy cluster cosmological simulations}

Numerical cosmological simulations of DM halos reproduce the properties of the large-scale structure in the Universe. In this work, we used two simulated clusters of galaxies for testing our orbital model and infer its uncertainties and systematics.

All simulations discussed below have been performed with the code \textsc{gadget} \citep{2005MNRAS.364.1105S}, and post-processed using the friends-of-friends (\textsc{fof}) algorithm with linking length $\ell=0.2$ times the mean interparticle separation \citep{1985ApJ...292..371D}. Substructures within \textsc{fof} groups have been subsequently identified with the \textsc{subfind} algorithm \citep{2001MNRAS.328..726S}. In the analysis, we considered the mass and centre of mass position and velocity of all the sub-haloes with at least 10 dark matter particles.

We initially evolved a periodic volume of size $L=500\,h^{-1}~\mathrm{Mpc}$ containing $512^3$ collisionless matter particles down to $z=0$. The initial conditions (ICs) were generated with \textsc{music} \citep{2011MNRAS.415.2101H} with second order Lagrangian perturbation corrections. We adopted a $\Lambda$CDM cosmology with parameters $(\Omega_0,\Omega_\Lambda,h,\sigma_8,n_{\rm s})=(0.285,0.715,0.695,0.828,0.9632)$ derived from the best-fitting model of WMAP9 and BOSS DR9 data of \cite{2013MNRAS.433.1202S}. The input linear power spectrum for the ICs generator was computed with \textsc{camb} \citep{2000ApJ...538..473L}, assuming the universal baryon fraction $\Omega_{\rm b}=0.046$. 

We selected two relatively relaxed clusters of mass $M_{200}=5.07\times 10^{14}~h^{-1} \mathrm{M}_\odot$ (CLUSTER 1) and  $M_{200}=6.55\times 10^{13}~h^{-1} \mathrm{M}_\odot$ (CLUSTER2) from the aforementioned cosmological volume, and re-simulated them with higher resolution by generating multi-level (zoom-in) ICs with \textsc{music}. The $r_{200}$ radius of the clusters turned out to be 1.30 $h^{-1}$ and 0.65 h$^{-1}$ Mpc for CLUSTER1 and CLUSTER2, respectively. The effective number of particles in the highest-resolution region is $8192^3$, corresponding to a mass resolution of $1.8\times10^7\,h^{-1}~\mathrm{M}_\odot$. The co-moving gravitational softening was set to $\epsilon=2.4\,h^{-1}~\mathrm{kpc}$, with its maximum physical value at $z=3$ ($\epsilon_{\rm phys}=0.6\,h^{-1}~\mathrm{kpc}$). The entire higher resolution region consists of $\sim 1.4\times 10^7$ particles.

From these simulations, we obtained the $\beta$ profile computed  in spherical shells as a function of the clustercentric distance. The resulting values of $\beta (r)$ are shown in Fig. \ref{betasimu0091} and \ref{betasimu3684}. The simulated clusters were used to solve Jeans equation with our code and obtain the velocity anisotropy profile. The clusters were projected along three different lines of sight. For each projection, we computed the projected $\Sigma(R)$ and $\sigma_{p}(R)$ and we fitted the mass distribution of the simulated clusters by a NFW density profile. Using our code we computed the $\beta (r)$ profiles from the projected simulated clusters and we compare them with the anisotropy profile in 3D shells in Figs \ref{betasimu0091} and \ref{betasimu3684}. The grey shaded area represent the maximum and minimum values of $\beta$ obtained at fixed radius from the three different projections of the clusters. Notice that the values of $\beta$ obtained from the spherical shells in the clusters are within the grey shaded area. 

For a spherically symmetric galaxy cluster we expect that the value of $\beta$ does not depend on the projection of the cluster. Therefore, the broad grey region observed in Fig. \ref{betasimu0091} is related to an elongated structure on the sky of this cluster which is not completely spherically symmetric. The other cluster, instead, presents a more spherical shape with a $\beta$ profile not strongly depending on the projection (see Fig. \ref{betasimu3684}).

We did not consider the effect of interlopers on the values of $\beta$.  The selection of the sub-halos of the clusters with a FOF algorithm makes the expected number of interlopers small, similarly to what happens with the observational data from A\ 85. For this cluster, the caustic method only returns 12 interlopers within $\pm 4 \sigma_{c}$, where $\sigma_{c}$ is the cluster velocity dispersion \citep[see ][]{agulli2016}. This number represents less than 3$\%$ of cluster members in A\ 85.


   \begin{figure}
   \centering
\includegraphics[width=\hsize]{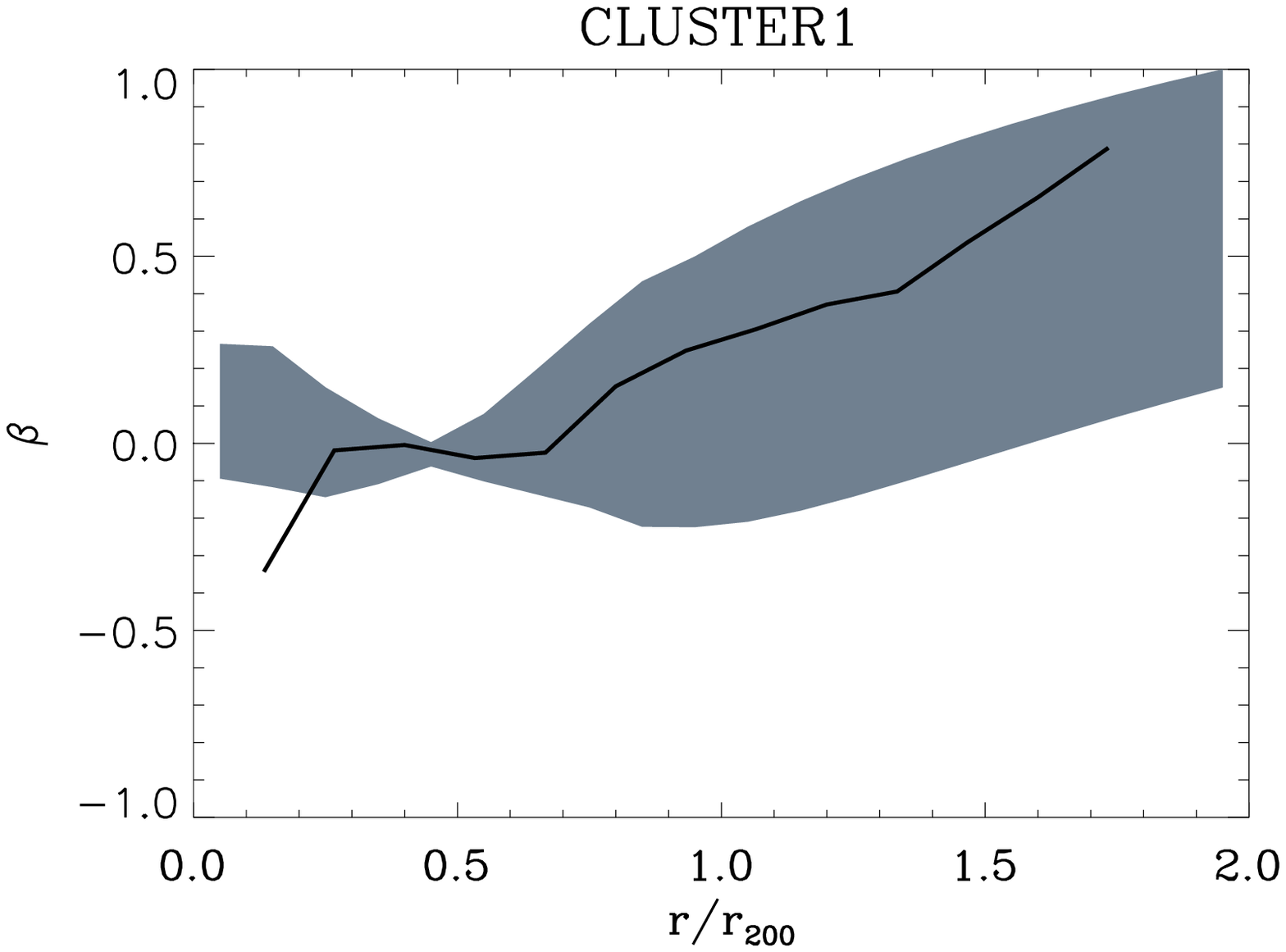}
   \caption{Velocity anisotropy profile measured in spherical shells (solid line) and obtained from the solution of the Jeans equation (grey shaded area) for the simulated CLUSTER1.}
             \label{betasimu0091}%
    \end{figure}

   \begin{figure}
   \centering
\includegraphics[width=\hsize]{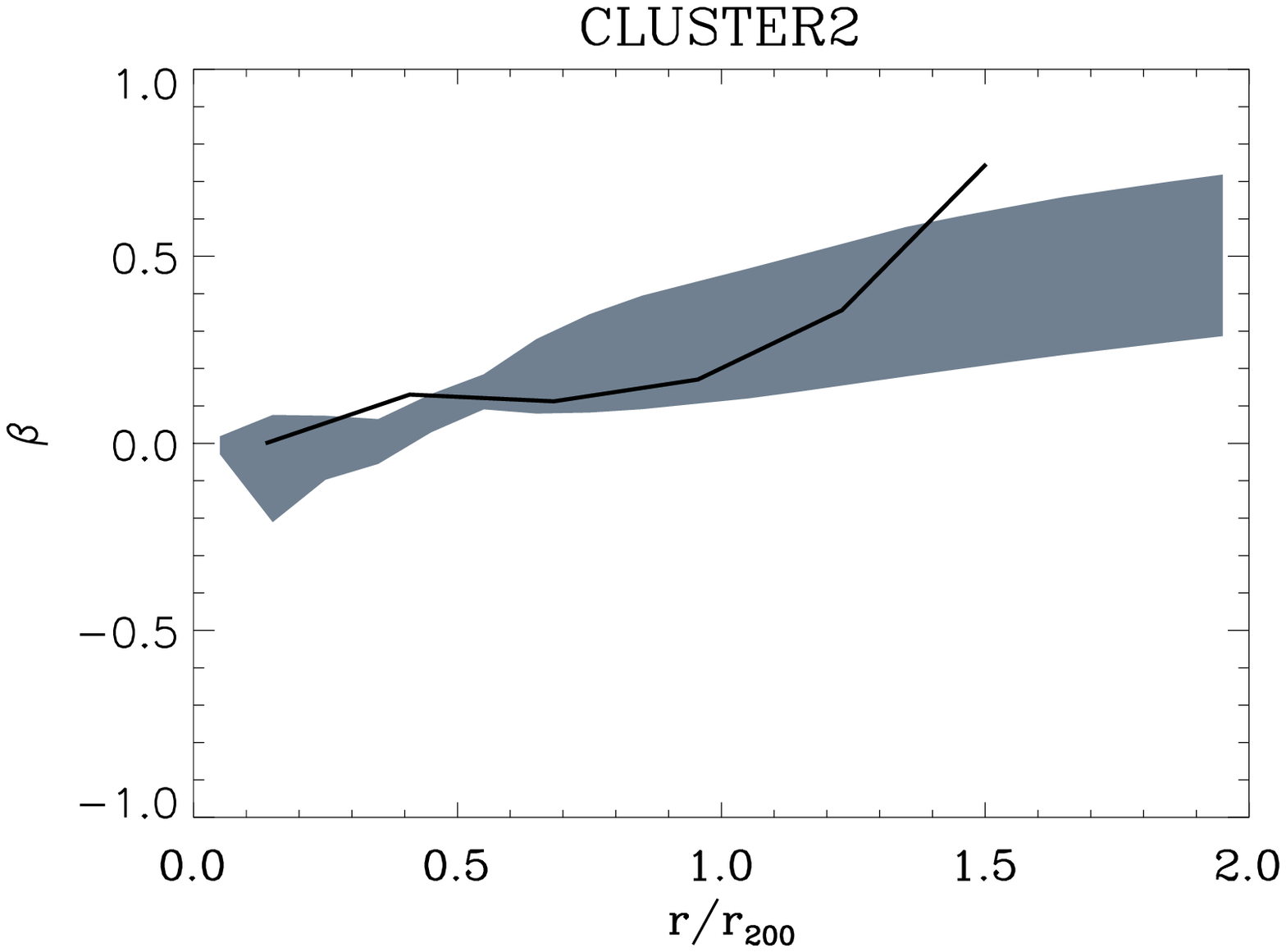}
   \caption{Velocity anisotropy profile measured in spherical shells (solid line) and obtained from the solution of the Jeans equation (grey shaded area) for the simulated CLUSTER2.}
             \label{betasimu3684}%
    \end{figure}

\section{Uncertainties on $\beta(r)$ as a function of the observed number of galaxies}

The velocity anisotropy is a statistical parameter which requires a large number of galaxies in order to have small uncertainties. We used one of the simulated clusters in order to check the uncertainties on $\beta$ as a function of the number of galaxy halos used for its computation. We selected the cluster with the largest number of  DM subhalos (CLUSTER1).

We analysed the effect of the number of members on $\beta (r)$, spanning the range between 50 and 500 halos. We used random samples of halos 
to compute the $\beta$ profile for the three different projections. Therefore, considering $\beta(r)$ the profile obtained from the Jeans equation and $\beta_{clus}$ the values for the different number of subhalos, we obtained the standard deviation $\beta(r) - \beta_\text{clus} (r)$ within $1.5 \, r_{200}$. We run 1000 Monte Carlo realisations for each adopted number of members. We present the values of $\beta_\text{clus}$ with their relative uncertainties depending on the number of subhalos considered in Fig. \ref{betauncer}. Moreover, we show the standard deviation as a function of the number of subhalos in Fig. \ref{betauncer2}.

Fig. \ref{betauncer2} shows that the uncertainty substantially drops around 100 subhalos. In particular, the uncertainty is smaller than 0.2 when 100 or more subhalos are considered, while when only 50 subhalos are used in the analysis the uncertainty is larger than 0.4.

   \begin{figure}
   \centering
\includegraphics[width=\hsize]{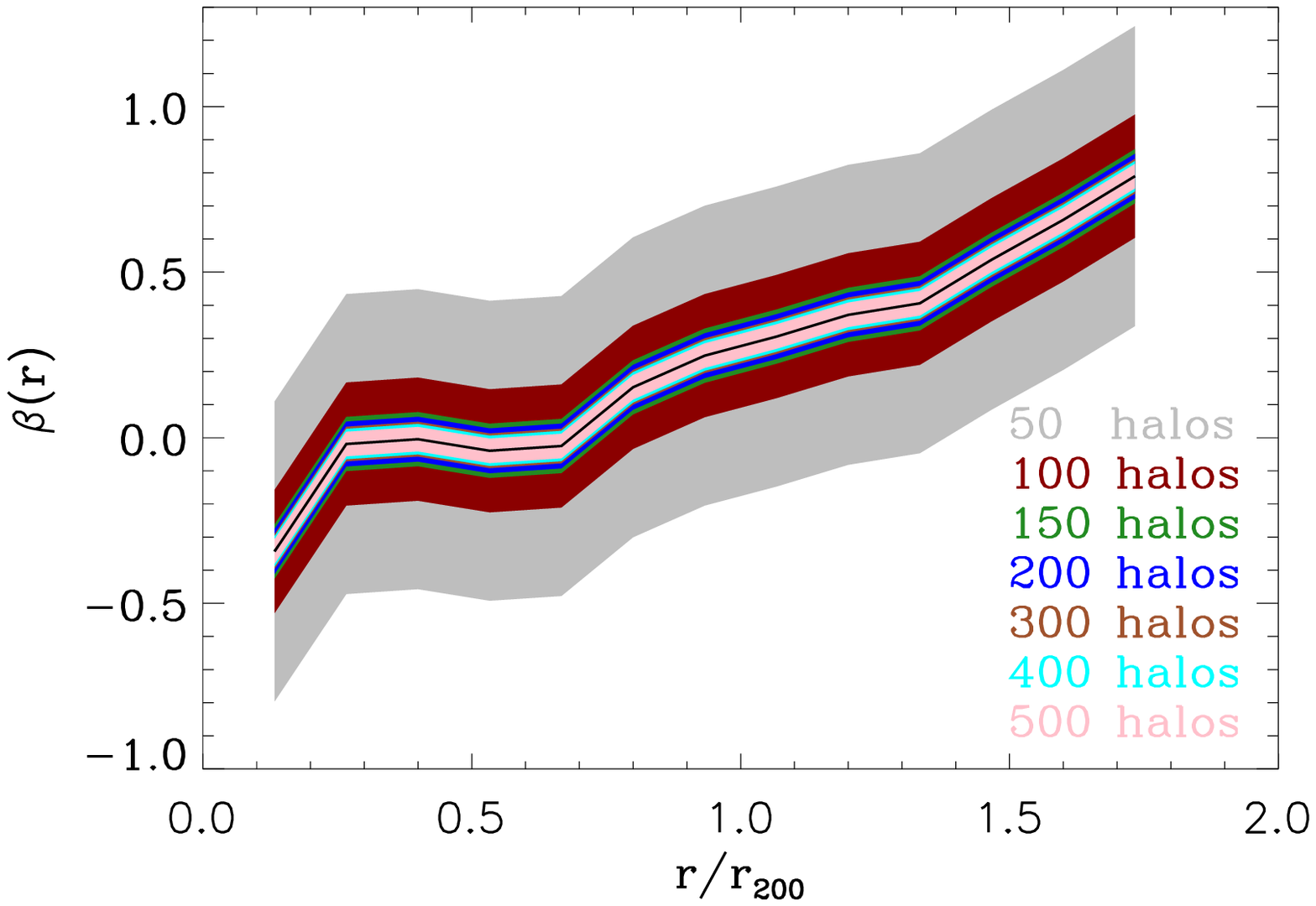}
   \caption{Anisotropy radial profile for a simulated cluster with 981 halos (Black line). The coloured shade regions represent the mean uncertainties on $\beta(r)$ for different numbers of halos (see text for more details).}
             \label{betauncer}%
    \end{figure}
    
       \begin{figure}
   \centering
\includegraphics[width=\hsize]{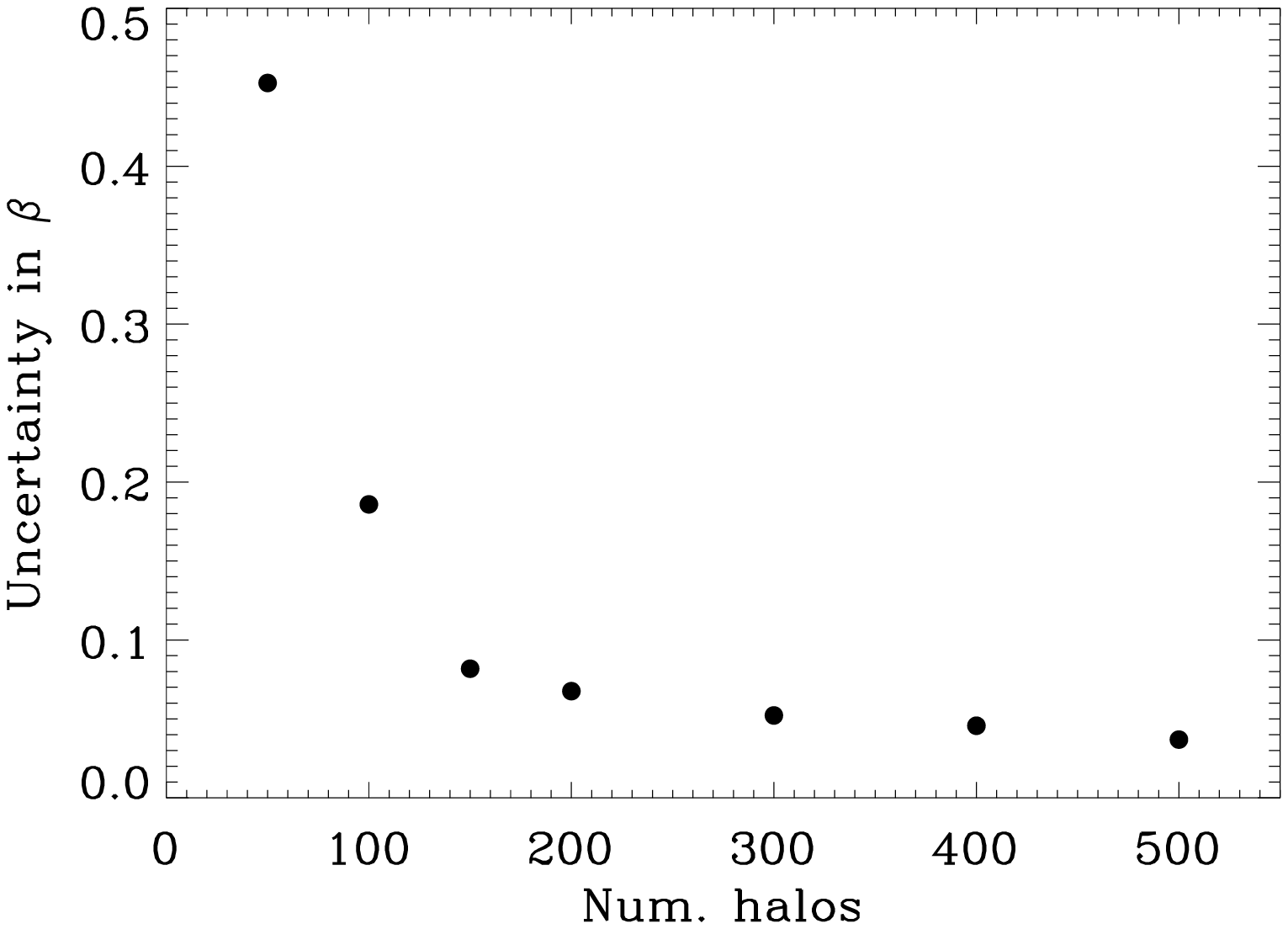}
   \caption{Standard deviation of $\beta - \beta_{clus}$ as a function of the number of halos considered for the solution of the Jeans equation.}
             \label{betauncer2}%
    \end{figure}
    
\section{Uncertainties on $\beta$ due to the velocity uncertainty}

Observational data are always affected by uncertainties. We analyse here how the uncertainties on the recessional velocities affect the anisotropy parameter. 

For A\,85, we have velocities from VIMOS@VLT, AF2@WHT, SDSS and NED. Since these data have different uncertainties on the redshift determination \citep[see][]{agulli2016}, we run Monte Carlo simulations to study their effects on the $\beta$ profile. In particular, we assigned uncertainties of 500 km/s to the members with VIMOS velocity information, 200 km/s for AF2 redshifts and 100 km/s for SDSS or NED data. We run 1000 realizations sampling the member velocities from a Gaussian distribution with width equal to their uncertainties. We used only the NFW mass profile by \cite{rines2006} for simplicity. 

We present the $\beta$ profile for the cluster and the maximum differences among the anisotropy values from the simulations at each radius in Fig. \ref{vel_err}. The uncertainty obtained on the anisotropy profile is smaller than 0.1 at every radius and it is mainly affected by the large dispersion for the VIMOS measurements. Therefore, the errors on A\,85 velocities do not significantly affect the results reported in this paper. 

       \begin{figure}
   \centering
\includegraphics[width=\hsize]{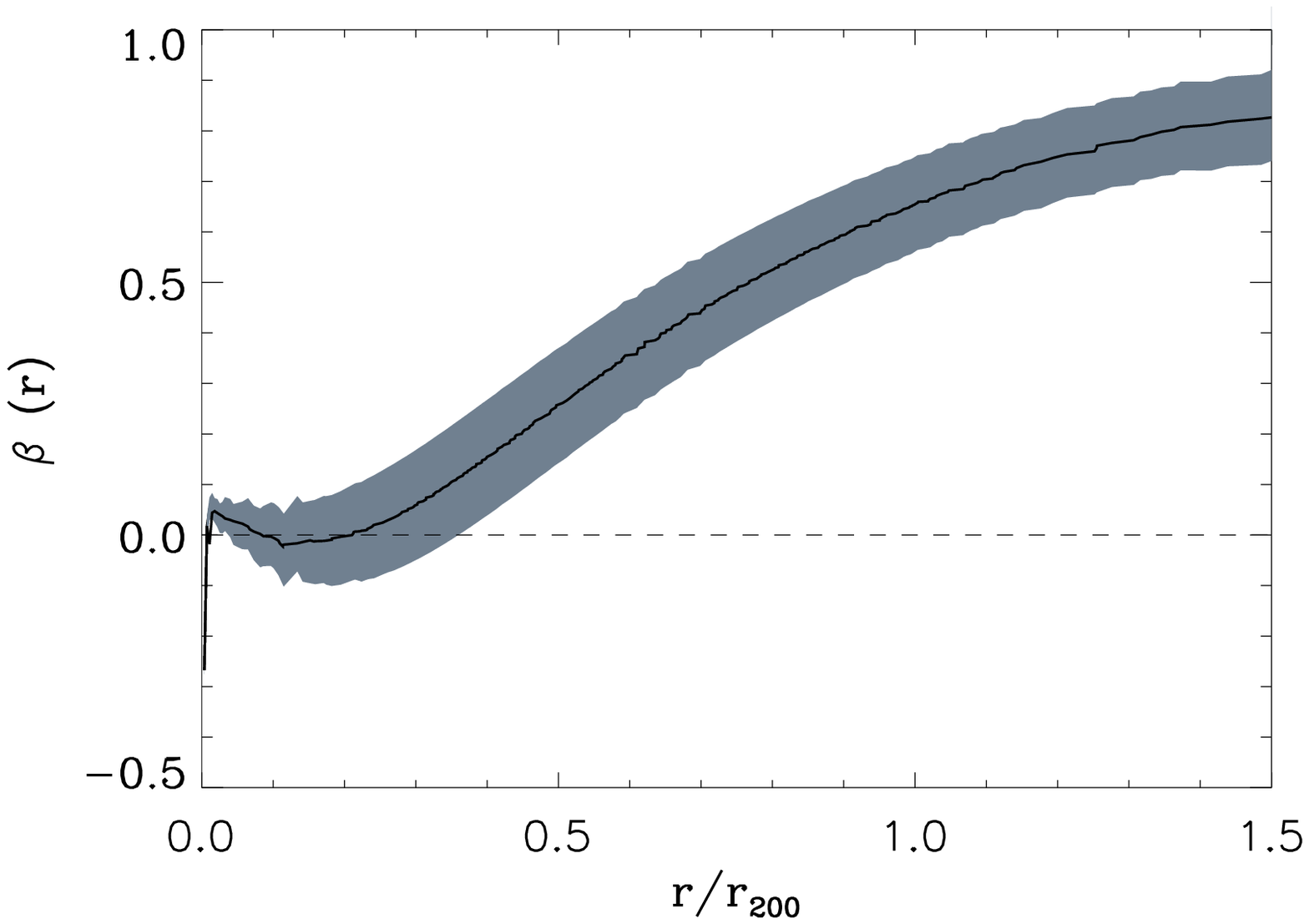}
   \caption{$\beta$ as a function of radius for one of the mass models of A\,85 \citep[][]{rines2006}. The shaded grey area represents the uncertainty on $\beta$ due to the velocity uncertainties.}
             \label{vel_err}%
    \end{figure}


\bsp	
\label{lastpage}
\end{document}